\documentclass[apj]{emulateapj}

\usepackage[english]{babel}

\shorttitle{Lensing \& reddening effects of MgII absorbers}
\shortauthors{M\'enard et al.}

\def\MgII{MgII}
\def\d{\mathrm{d}}
\def\be{\begin{equation}}
\def\ee{\end{equation}}
\def\bea{\begin{eqnarray}}
\def\eea{\end{eqnarray}}

\def\N{\mathrm{N}}

\begin{document}

\title{Lensing, reddening and extinction effects\\ of MgII absorbers
from $z=0.4$ to $z=2$}

\author{Brice M\'enard         \altaffilmark{1},
  Daniel Nestor           \altaffilmark{2},
  David Turnshek          \altaffilmark{3}, 
  Anna Quider \altaffilmark{3},\\
  Gordon Richards \altaffilmark{4},
  Doron Chelouche \altaffilmark{5}
\& Sandhya Rao \altaffilmark{3}}

\altaffiltext{1}{Canadian Institute for Theoretical Astrophysics, 
University of Toronto, 50 St George street, Toronto ON}

\altaffiltext{2}{Institute of Astronomy, University of Cambridge, 
Madingley Road, Cambridge. CB3 0HA, U.K.}

\altaffiltext{3}{Dept. of Physics and Astronomy, University of Pittsburgh, 
Pittsburgh, PA 15260, USA} 

\altaffiltext{4}{Department of Physics, Drexel University, 
3141 Chestnut Street, Philadelphia, PA 19104.}

\altaffiltext{5}{Institute for Advanced Study, Einstein Drive, 
Princeton NJ 08540, USA}

\begin{abstract}
  Using a sample of almost $7000$ strong MgII absorbers with
  $0.4<z<2.2$ detected in the SDSS DR4 dataset, we investigate the
  gravitational lensing and dust extinction effects they induce on
  background quasars.  After carefully quantifying several selection
  biases, we isolate the reddening effects as a function of redshift
  and absorber rest equivalent width, $W_0$. We find the amount of
  dust to increase with cosmic time as
  $\tau(z)\propto(1+z)^{-1.1\pm0.4}$, following the evolution of
  cosmic star density or integrated star formation rate.  We measure
  the reddening effects over a factor $30$ in E(B--V) and we find that
  $\tau\propto (W_0)^{1.9\pm0.1}$, providing us with an important
  scaling for theoretical modeling of metal absorbers. We also measure
  the dust-to-metals ratio and find it similar to that of the
  Milky Way.  In contrast to previous studies, we do not detect any
  gravitational magnification by MgII systems.  We measure the upper
  limit $\mathrm{\mu}<1.10$ and discuss the origin of the discrepancy.
  Finally, we estimate the fraction of absorbers missed due to
  extinction effects and show that it rises from 1 to 50\% in the
  range $1<W_0<6$~\AA.  We parametrize this effect and provide a
  correction for recovering the intrinsic $\partial N/\partial W_0$
  distribution.
\end{abstract}


\section{Introduction}

Quasar absorption lines provide us with a unique tool to probe the gas
content in the Universe. They offer an unmatched sensitivity up to
high redshift, allow us to detect arbitrarily faint objects, and
constrain the gas distribution in velocity space.  However, the lack
of direct spatial information tends to be a limitation for
understanding the nature of these systems and fundamental questions
still remain. One way to overcome this problem is to relate the gas
distribution to that of other components such as dark matter, stars
and dust by measuring their cross-correlations.

Galactic environments have been studied using various types of
absorption lines.  Among metal lines, the \MgII\ doublet,
$\lambda\lambda2796,2803\,$\AA, has been extensively used due to the
strength and wavelength of the transitions making it easily detectable
from the ground at $0.4\lesssim z\lesssim 2.2$
(\cite{Lanzetta+87,Steidel_Sargent92,Nestor+05,Prochter+06}).
MgII turns out to be a sensitive tracer of galactic environment: it
arises in gas spanning more than five decades of neutral hydrogen
column density (e.g. \citealt{Churchill+00}), and it has been shown
that the detection of a strong MgII absorber indicates the presence of
a nearby $\sim 0.1-10\,L_\star$ galaxy along the line-of-sight
(\citealt{Bergeron91,Steidel+94,Bergeron+92,
LeBrun+93,Steidel+94,Steidel+97}), and at low redshift \cite{Rao+06} have
shown that strong MgII absorbers are good tracers of Damped Lyman
alpha systems. 
For reviews of the progress on the determination of
these systems, see \citet{Tripp_Bowen05} and \citet{Churchill05}.

Recently, the size and homogeneity the SDSS has allowed several
statistical analyzes to be done: \cite{Bouche+06} measured the
cross-correlation between $z\sim0.5$ luminous red galaxies and MgII
absorbers. \cite{Zibetti+05,Zibetti+06} investigated the
distribution of absorber impact parameters and the emission properties
of MgII related galaxies.  They have shown that a cross-correlation
between light and gas can be detected up to 200 kpc and that stronger
absorbers are related to bluer galaxies. The SDSS has also provided us
with closely separated quasar pairs allowing us to analyze the
spectrum of background quasars to probe the gas distribution around
foreground ones \citep{Bowen+06,Bowen+07}. Such a technique allows us
to compare the gaseous halos of quasars and galaxies and it can
ultimately be used to test the quasar unification scheme on large
scales \citep{Chelouche+07}.

However, despite significant observational progress fundamental
questions such as the nature of the gas seen in absorption and its
spatial distribution still remain. More importantly, we still do not
understand the process(es) responsible for the observed values of MgII
rest equivalent widths. As strong MgII absorbers are known to be
tracers of galaxies, they are expected to induce gravitational lensing
and reddening effects to the light of their background
quasars. Correlations between quasar magnitudes and the presence or
strength of metal lines can therefore be used in order to quantify the
relationships between gas, dark matter and dust. Such an approach also
allows us to quantify the extinction and magnification biases in
samples of optically selected quasars.

\subsection{Dust reddening}

The presence of dust associated with various types of absorbers has
been reported by many authors: \cite{Pei91} \& \cite{Pei92} found
quasars with DLAs to be on average redder than quasars
without. \cite{Richards01} observed a similar effect with CIV
absorption lines.  Using the 2dF survey \cite{mp03} found evidence
of reddening due to strong MgII systems. \cite{Hopkins04} showed that
a fraction of SDSS quasars are reddened with SMC-type dust and argued
that this effect comes predominantly from dust located at the quasar
redshifts.  Using the SDSS database, \cite{Wild+05,Wild06} found an
E(B-V)$\simeq0.06$ associated with CaII absorbers, and \cite{York+06} \&
\cite{Khare+05} measured the mean reddening effects of about 800
quasars with $z>1$ MgII absorbers.
Constraints on dust can also be obtained by measuring the relative
abundances of volatile and refractory elements in order to infer a
depletion level
(e.g. \citealt{2005A&A...444..461V,vladilo+06,Nestor+03}).

While the presence of dust reddening associated with metal absorbers
has been convincingly shown, its detailed properties (dependencies on
redshift, rest equivalent width, etc.) need to be quantified in order
to understand the underlying physics and connect theory to
observations.

The presence of dust along given lines-of-sight also implies the
existence of a dust obscuration bias that can affect quasar and
absorber number counts. Using 75 radio-selected quasars,
\citet{Ellison+04} investigated such an effect and did not find any
difference (at the 20\% level) between the incidence of MgII absorbers
in radio-selected and optically-selected quasars. This result shows
that \emph{in general} optically-selected quasars are not strongly
affected by an extinction bias, but it does not provide us with the
full information: it does not show how the shape of the
observed distribution of absorber rest equivalent widths is affected
by dust extinction.

In order to unveil the relationships between extinction and absorption
properties, we present a new analysis based on the Sloan Digital Sky
Survey (SDSS, \citealt{York00}) using the fourth data release (DR4).
After having quantified the efficiency of the SDSS pipeline to detect
reddened quasars, we use a sample of about 7,000 MgII absorbers --
i.e. almost an order of magnitude larger than previous studies
-- to measure the excess reddening induced by these systems and
investigate its dependence on absorption rest equivalent width and
redshift.

\subsection{Gravitational lensing}
\label{section_lensing}

In addition to cause reddening effects, the presence of a galaxy along
the line-of-sight is expected to give rise to gravitational lensing
effects.  Several cases of strong gravitational lensing of quasars due
to an absorbing galaxy are known (e.g. \citealt{Turnshek97}). However,
in general, the impact parameters of MgII absorbers are larger than a
few kpc and gravitational lensing effects occur in the weak regime and
solely change quasar magnitudes
(\citealt{Bartelmann96,Perna97,Smette97,Bartelmann98,Menard05}). The
detection of such an effect would ultimately allow us to constrain the
mass of MgII systems.

Several authors have attempted to detect the statistical magnification
induced by MgII systems. Some have looked for redshift distribution
changes (\citealt{Thomas90,Borgeest93}) but did not find any
compelling evidence for gravitational lensing.  Others have
investigated the occurrence of metal absorbers (CIV, SiIV, MgII) on
bright and faint quasars (\citealt{York91,vandenberk96,Richards99})
and reported possible magnification effects for certain absorber
species only.  \citet{mp03} used the 2dF-Quasar survey to compare the
number of quasars with and without MgII absorbers as a function of
magnitude and reported a relative excess of optically bright
radio-selected quasars with absorbers as well as indications of
reddening effects. Murphy \& Liske (2004) applied the same method to a
sample of Damped Lyman-$\alpha$ systems and reported a similar trend
found at 2$\sigma$. Using radio-selected quasars, \citet{Ellison+04}
reported a possible excess of optically bright quasars with MgII
absorbers, and more recently \cite{Prochaska+05} found that quasars
with high-redshift DLAs are on average brighter than quasars
without. The various results found among all these analyzes turn out
to be rather difficult to combine because the absorber species,
redshift and equivalent width ranges differ. In addition, as noted in
some of these analyzes, biases due to selection effects can mimic the
lensing signal: it is easier to detect absorption lines in high
quality (or high S/N) spectra, however, these spectra tend to
correspond to bright objects. This selection bias results in a
preference for detecting absorption lines in brighter quasars, which
is an effect similar to the signal of interest. Unfortunately, none of
these works could convincingly demonstrate that the level of such
systematics was sufficiently low in their analysis, leaving the
amplitude of lensing effects rather uncertain.

In this paper we attempt to make a significant improvement in making
such a measurement by overcoming the problem of systematic effects
using Monte Carlo simulations and by reaching a much higher precision
thanks to the size and homogeneity of the SDSS. We use a robust
semi-automatic absorption line finder algorithm, carefully select a
population of reference quasars, test potential biases, and use Monte
Carlo simulations to measure and quantify the level of systematics of
our selection procedure. To detect a magnification signal we will
apply the method proposed by \cite{Menard05} and look for a
correlation between the presence of absorbers and a change in the mean
magnitude of their background quasars. Contrarily to previous studies,
this approach allows the use of a non-parametric estimator.

The outline of the paper is as follows: in \S\ref{section_formalism}
we introduce the formalism of dust extinction and gravitational
magnification. We describe the data used in our analysis in
\S\ref{section_data} as well as the selection of the different samples
of quasars. The magnitude shifts and color changes due to the presence
of absorbers along quasar lines-of-sight are presented in
\S\ref{section_results}. We discuss the results and conclude in
\S\ref{section_conclusion}.

\section{FORMALISM}
\label{section_formalism}

Strong MgII absorbers (usually defined with $W_0(2796)>0.3$ \AA) are
known to be tracers of galaxies with luminosities ranging from 0.1 to
several $L_\star$.  The presence of such a system seen along the
line-of-sight of a background source can therefore modify its
brightness in two ways: first, the presence of dust in and/or around
the absorber can extinct and redden the source's light and second, it
can act as a gravitational lens and amplify the flux of the
source. This can be described by
\begin{equation}
\mathrm{f}(\lambda)=\mathrm{f}_\mathrm{ref}(\lambda)\,
\mu\,e^{-\tau_\lambda}\,,
\end{equation}
where $\mathrm{f}(\lambda)$ is the observed flux of a source behind an
absorber, $\mathrm{f}_\mathrm{ref}(\lambda)$ is the flux that would be
observed without intervening system, $\tau_\lambda$ is the optical
depth for extinction induced by the galaxy and $\mu$ the gravitational
amplification.  The corresponding magnitude change can then be written
\bea
\delta m &=& m-m_{\mathrm{ref}}\nonumber\\
&=&-2.5\,\log(\mu)+\frac{2.5}{\ln 10}\,\tau_\lambda\;.
\label{eq:2}
\eea 
These two contributions have different wavelength and redshift
contributions. One can therefore attempt to constrain them separately.

\subsection{Measuring reddening effects}
\label{section_measured_reddening}

Following \cite{Menard05}, we introduce the observed mean magnitude
difference between a population of quasars with absorbers and one
without:
\begin{equation}
\Delta\langle m_j \rangle = \langle m_j \rangle -\langle m_{j,ref}
\rangle\,,
\label{eq_magnitude_shift}
\end{equation}
where $m_j$ is the magnitude in a given band $j$.  In the next section
we will measure this quantity using a large sample of MgII absorbers
from the SDSS. We will then investigate how $\Delta\langle m_j
\rangle$ varies as a function of absorber rest equivalent width and
redshift.  Similarly, we can then define the mean excess color
\begin{equation}
\langle E(j-k) \rangle= \Delta\langle m_j \rangle - \Delta\langle m_k
\rangle
\end{equation}
where $j$ and $k$ denote two different bands. It should be noted that
such a quantity is not sensitive to gravitational lensing effects as
it no longer depends on the magnification $\mu$.  It can be used to
probe the statistical properties of the extinction curve of absorbers
systems as well as the dust column densities as a function of absorber
rest equivalent width.\\

In optical surveys, sources like quasars are both magnitude and color
selected.  When dust extinction occurs, such sources can become
undetectable because they become too faint and/or too red.
As long as sources are no longer detectable because of a limiting
magnitude and not because of a color cut, we have
\begin{equation}
\langle E(j-k) \rangle=
\frac{2.5}{\ln 10} \langle \tau(\lambda_j) - \tau(\lambda_k) \rangle\,.
\end{equation}
However, if the sources become undetectable because they no longer
satisfy a color cut, the measured reddening is then lower than the
intrinsic one. Such a property depends on the details of the quasar
selection procedure and will be addressed in section
\ref{section_reddening_bias}.\\

The current study is based on the photometry of SDSS quasars using the
${u,g,r,i,z}$ filter set. However, historically people have quantified
color excesses using the Johnson $B$ and $V$ filters. In order to be
able to present our results in similar units, we provide here a list of
useful conversions.
First, it should be noted that previous statistical studies
(\citealt{Khare+05,Menard+05,York+06,Wild06}) have indicated that, on
average, the extinction curves of MgII selected galaxies are
consistent with that of the SMC, i.e. do not present an excess around
the rest frame wavelength of 0.2$\micron$ (in the optical, such a
feature can only be detected for systems at redshift greater than
$\sim1$ and is therefore irrelevant for lower redshift galaxies).  In
this paper we will therefore consider only SMC-type extinction curves.
We will also see in section \S\ref{section_results} that such a
property is confirmed by the present analysis.

In the range 0.1 to 1 $\micron$, the SMC extinction \-cur\-ve
\footnote{Templates of the reddening curves can be obtained from
  http://www.astro.princeton.edu/$\sim$draine}
can be very well approximated by a power law
$A_V\propto\lambda^{-1.2}$ \citep{Prevot+84}.  By using the $g$ and
$i$ filters, the conversion between the two color systems therefore
reads
\begin{eqnarray}
E(g-i)&=& \frac{\lambda_g^{-1.2}-\lambda_i^{-1.2}}
{\lambda_B^{-1.2}-\lambda_V^{-1.2}}\times E(B-V)\nonumber\\
& \simeq& 1.55 \times E(B-V)
\end{eqnarray}
and the average wavelengths for a flat spectrum are given by
$\{\lambda_g,\lambda_i,$ $\lambda_B,\lambda_V\} =\{468,748,435,555\}$
nm. Using the effective wavelength for realistic quasar spectra does
not change the results significantly.  In practice, two frames are
interesting for considering colors excesses: in order to quantify the
fraction of sources missed because of an extinction bias,
\emph{observed} E(B-V) values are relevant. However, for estimating
the amount of dust associated with an absorber system the \emph{rest
frame} E(B-V) is the quantity of interest. For an SMC-like extinction
curve we have
\begin{equation}
E(B-V)_{obs}=E(B-V)_{rest}\,(1+z)^{1.2}\,.
\label{eq_expected_z_dependence}
\end{equation}
In a given band, we will therefore observe almost four times more
extinction by an absorber at $z=2$ than the same system located at
$z=0$.  For an SMC extinction curve we also have $A_v\simeq C_A \times
\mathrm{E(B-V)}$ where the proportionality coefficient $C_A$ depends on
the redshift of the dust and varies from 3.2 to 2.0 when $z$ is in the
range 0.5 to 2.0, with a mean value of 2.6.

\subsection{Measuring gravitational lensing effects}
\label{section_lensing}

In this section we show how magnification effects can (or cannot)
affect the observed magnitude of distant sources like quasars.  For
more details we refer the reader to the original discussion presented
in \cite{Menard05}.

Here we will first assume that extinction effects can be neglected
(which is expected at sufficiently large wavelengths). Let us
consider an area of the sky which is large enough for the mean
magnification to be close to unity. In this area, let us consider a
population of sources, with a fraction $f$ of them lying behind an
absorber system. We will write the magnitude distribution of the
sources behind absorbers $\mathrm{N}(m)$, and $\mathrm{N_{ref}}(m)$
for the non-absorbed sources.  We then have the following relation
between the magnitude number counts:
\begin{equation}
\N(m)=f\times\int \N_\mathrm{ref}(m-\delta m)
\;\mathrm{P}(\delta m) \; \d(\delta m)\,,
\label{eq_convolution}
\end{equation}
where $\delta m$ is the induced magnitude shift and
$\mathrm{P}(\delta m)$ is the probability density to have a given
value of $\delta m$ along a quasar line-of-sight.  Note that in the
case of absorber systems, $f$ is not known. Observational estimates of
the number density of absorbers are necessarily affected by lensing
and/or extinction effects due to the need of background quasars to
observe absorber systems. Therefore the only usable information from
the previous equation comes from the shape of the distribution
$\N(m)$, i.e. its moments. In this paper we will focus only on the
first moment, i.e. we will attempt to constrain the mean magnification
and extinction effects by measuring the mean magnitude shift induced
by the presence of absorber systems.

By defining $\langle m \rangle$ and $\langle m \rangle_{ref}$ to be
the observed mean magnitudes of the two quasar populations, we see
that the \emph{observable} mean magnitude shift
$\langle\Delta{m_\mathrm{obs}}\rangle$ (eq. \ref{eq_magnitude_shift})
can be used to probe the unknown distribution $\mathrm{P}(\delta m)$.
It is important to note that the \emph{observed}-magnitude changes
$\langle\Delta{m_\mathrm{obs}}\rangle$ depend on both the
magnification effects \emph{and} the slope of the source luminosity
distribution:
\begin{equation}
\langle\Delta{m_\mathrm{obs}}\rangle=F[N(m)]\times \delta m\,,
\end{equation}
and in general we have
$\langle\Delta{m_\mathrm{obs}}\rangle<\langle\delta m\rangle$.  For
SDSS quasars with $i<20$, \cite{Menard05} has shown that we have
$\langle\Delta{m_\mathrm{obs}}\rangle\simeq 0.30\,\delta m$ where
$\delta m$ is an achromatic magnitude change. For a limiting magnitude
of $i=19.1$, we find $\langle\Delta{m_\mathrm{obs}}\rangle\simeq
0.25\,\delta m$.\\ It should be noted that if the number of sources as
a function of luminosity follows a power law then $F=0$ and magnitude
changes cannot be observed whatever the value of the induced magnitude
shift $\delta m$.
This absence of observable lensing effects can be understood in the
following way: gravitational magnification makes each source brighter
but allows also new sources to be detectable as they become brighter
than the limiting magnitude of a given survey. This latter effect
increases the number of faint sources and tends to diminish the mean
flux of the detected objects.  For a power law luminosity
distribution, the flux increase due to magnification is
exactly canceled by the additional faint sources that enter the sample
due to magnification, and lensing effects due to absorbers cannot be
observed.  Therefore, it is possible to detect changes in the
magnitude distribution of a population lensed by absorbers only if its
unlensed number counts depart from a power law as a function of
luminosity.  Extinction effects will be affected in a similar way
but the ability to observe sources in different bands allows us to
measure reddening effects and then infer the related extinction.

\section{DATA \& SELECTION BIASES}
\label{section_data}

\subsection{The quasar sample}

In order to isolate the magnitude changes due to the presence of
intervening absorber systems we need to compare the magnitude of
quasars with absorption lines to that of a reference quasar
population. Defining such a reference population without introducing
any selection bias is a difficult task. Indeed, the detection of
absorption lines depends on quasar redshift, spectrum S/N, and the
extinction and reddening effects they might induce on their background source.
All these effects have to be taken into account in
order to carefully select a population of reference quasars.

The first requirement is to work with a well defined sample of
quasars.  Among the large sample of quasars available from the SDSS
DR4 sample ($N\sim$50,000), we will use the primary spectroscopic
quasar sample, i.e. quasars with $i<19.1$ ($N\sim$29,000), as the
selection of these objects is based on a well defined series of color
cuts \citep{Schneider02}. We will not make use of fainter quasars
(observed with SDSS serendipity fibers) as they are not selected from
well defined selection criteria. Before characterizing the absorber
systems used in this study, the first step of our analysis will be to
investigate the sensitivity of the SDSS quasar target algorithm to
reddening and extinction.

\subsection{Detection of reddened quasars}
\label{section_reddening_bias}

\begin{figure*}[t]
\begin{center}
  \includegraphics[height=10cm,width=.9\hsize]{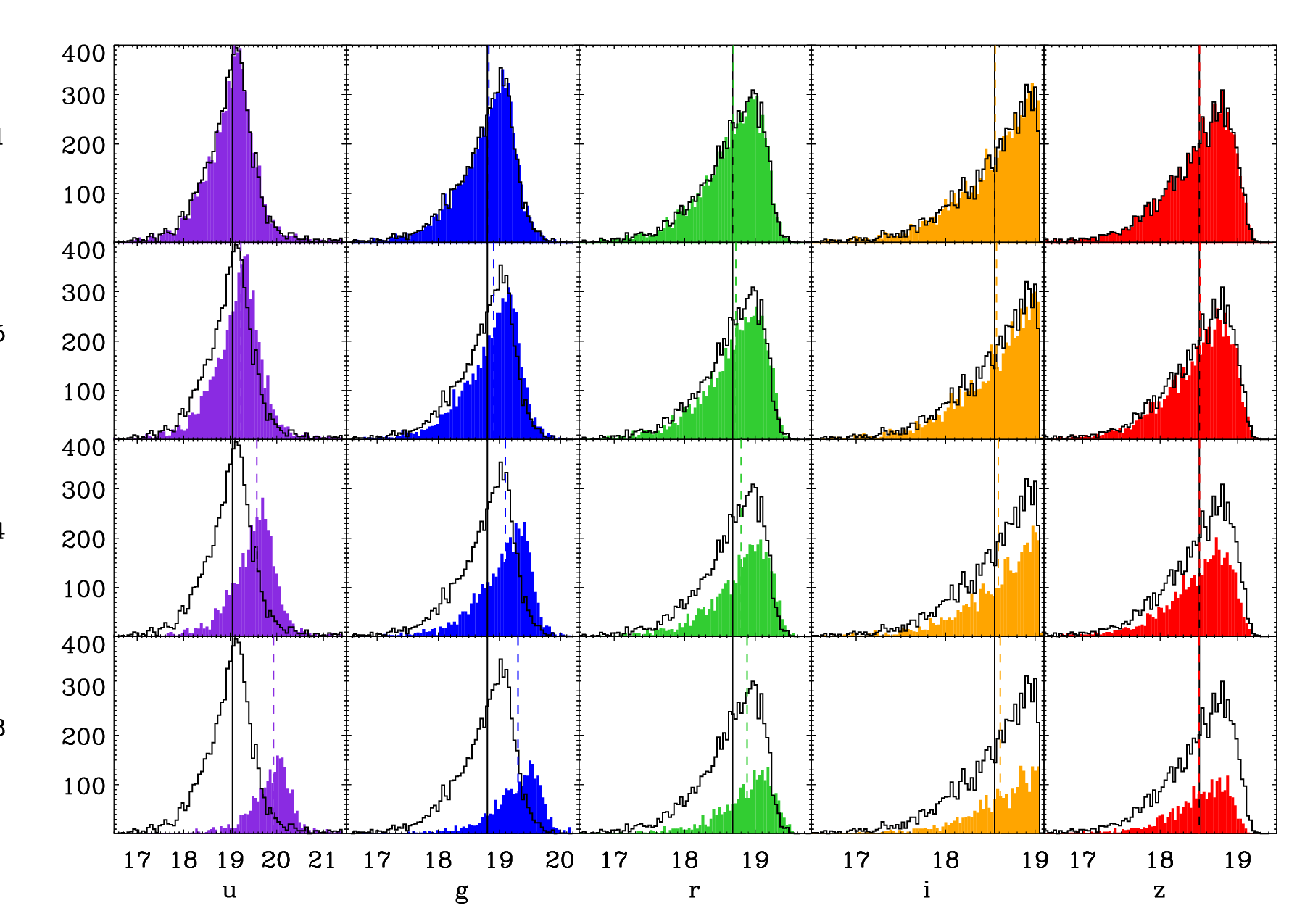}
  \caption{SDSS quasar detection sensitivity as a function of
  reddening. The solid-line histograms show the magnitude
  distribution of SDSS quasars without strong MgII absorber. The
  filled histograms show the fraction of objects that remain
  detectable after reddening effects. The vertical lines show the mean
  magnitude of each sample.}
  \label{plot_magnitude_grid}
\vspace{1cm}
  \includegraphics[width=.49\hsize]{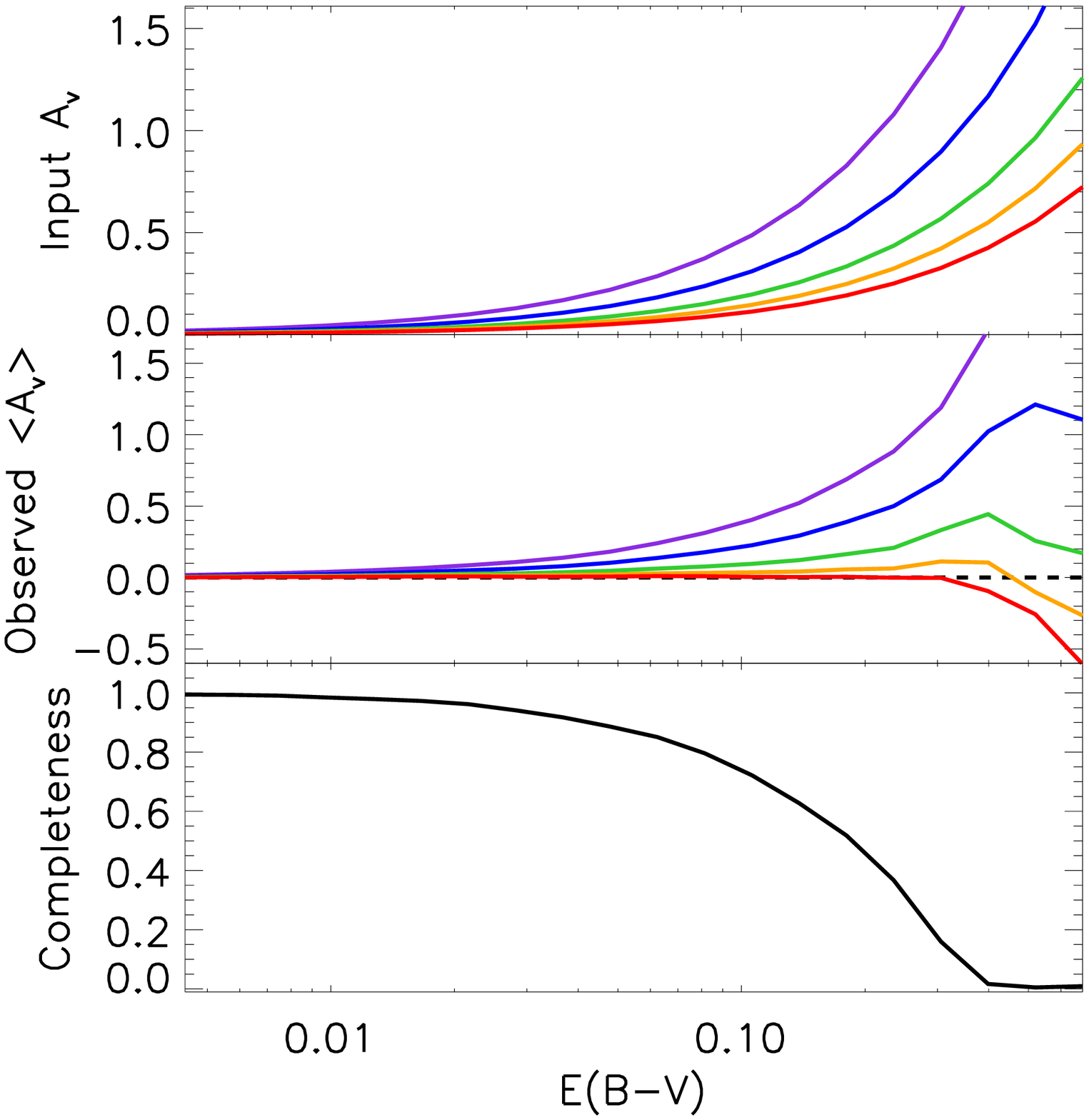}
  \includegraphics[width=.49\hsize]{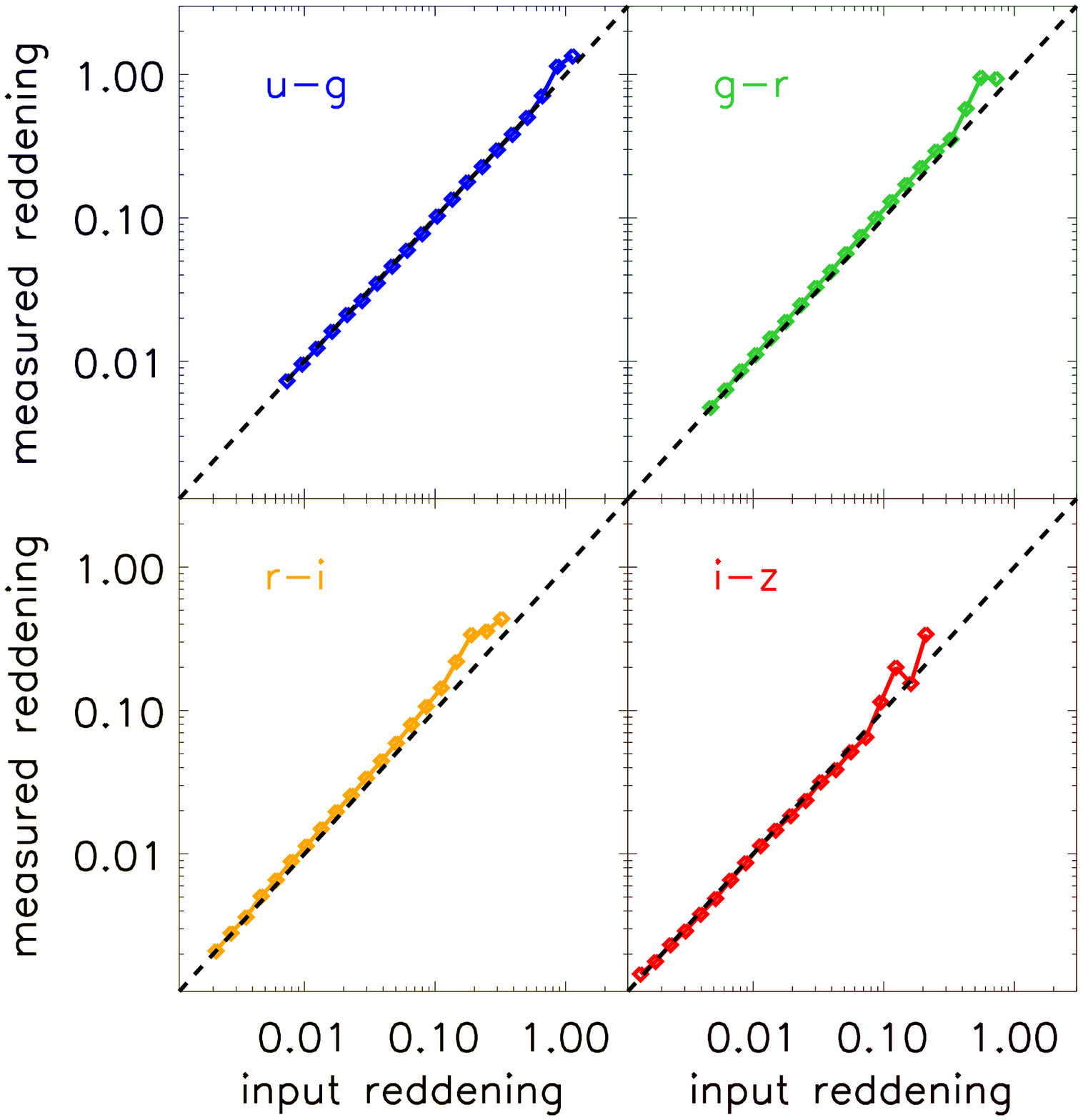}
  \caption{\emph{(Left)} Top panel: amount of extinction as a function
  of $E(B-V)$ for an SMC reddening curve. Middle panel: measured mean
  magnitude shift of the detected objects. We observe that the
  measured extinction is lower than the initially induced one. Note
  that the mean $z$ band magnitudes are not significantly affected by
  extinction.  Lower panel: fraction of SDSS quasars still detectable
  after having applied a given extinction. Similarly to the analysis
  done by \citet{richards+03}, the SDSS quasar algorithm is able to
  recover most quasars reddened by $E(B-B)<0.1$.}
  \label{plot_Gordon_mag_shift}
  \caption{\emph{(Right)} Measured reddening as a function of induced
  reddening. Even if the measured extinction differs from the induced
  ones, the color changes are not affected by the corresponding
  selection effects and in the reddening range of interest for the
  present study, the observed color changes of quasars can directly be
  used in order to quantify the amount of induced reddening by
  intervening absorbers. }
\end{center}
\vspace{1cm}
\end{figure*}

\begin{figure*}[t]
\begin{center}
  \includegraphics[height=6cm,width=.3\hsize]{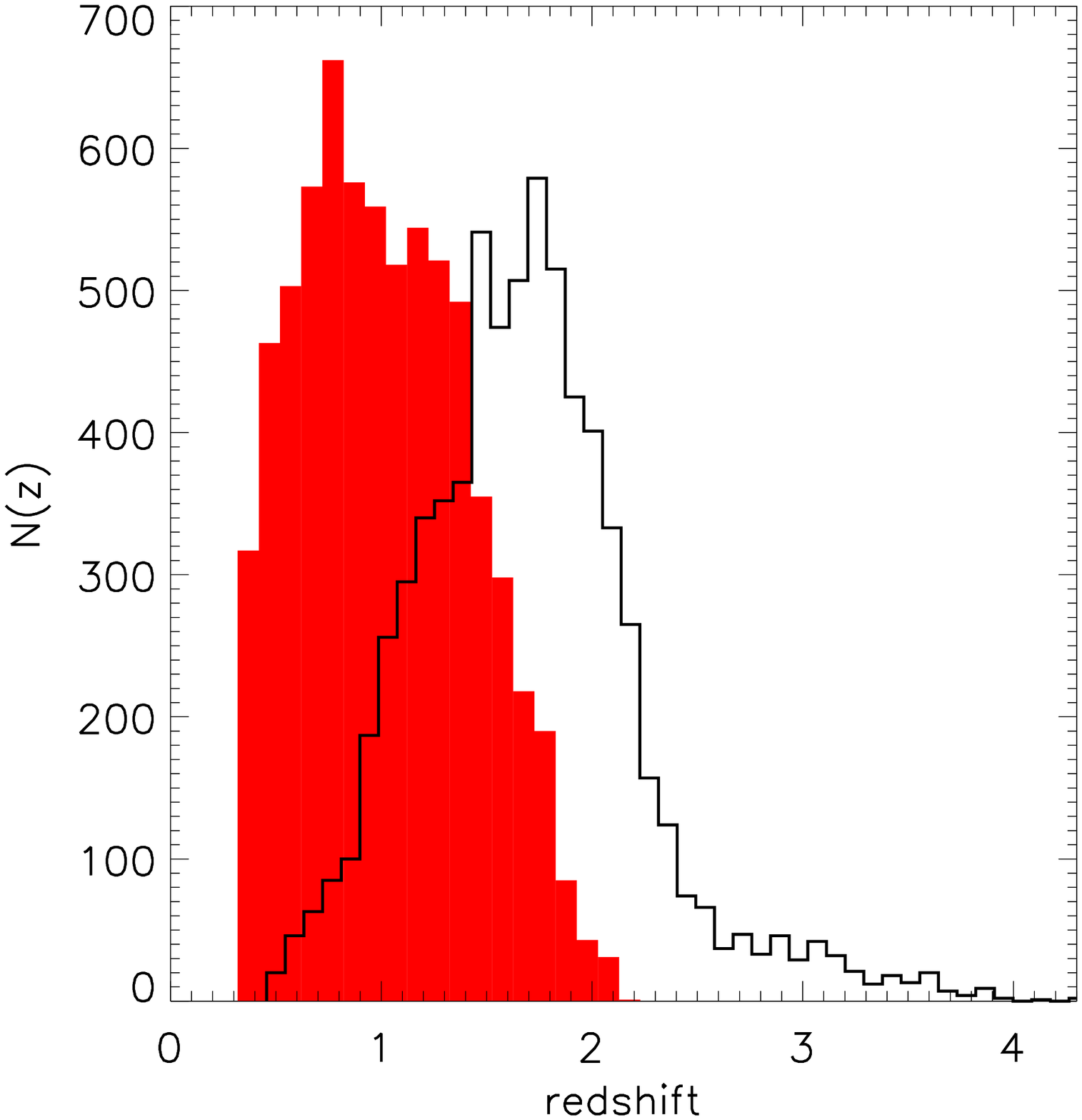}
  \includegraphics[height=6cm,width=.3\hsize]{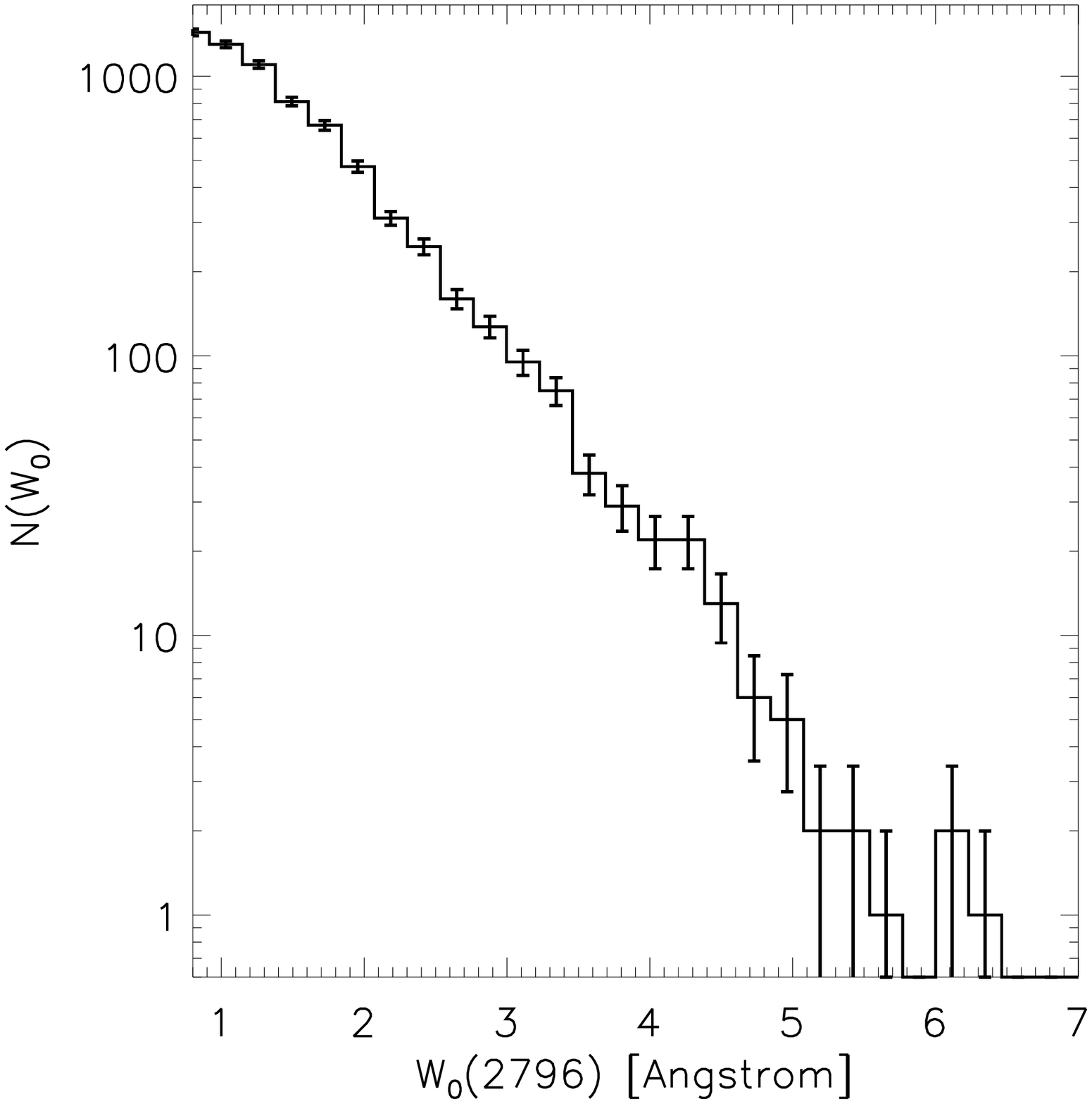}
  \includegraphics[height=6cm,width=.3\hsize]{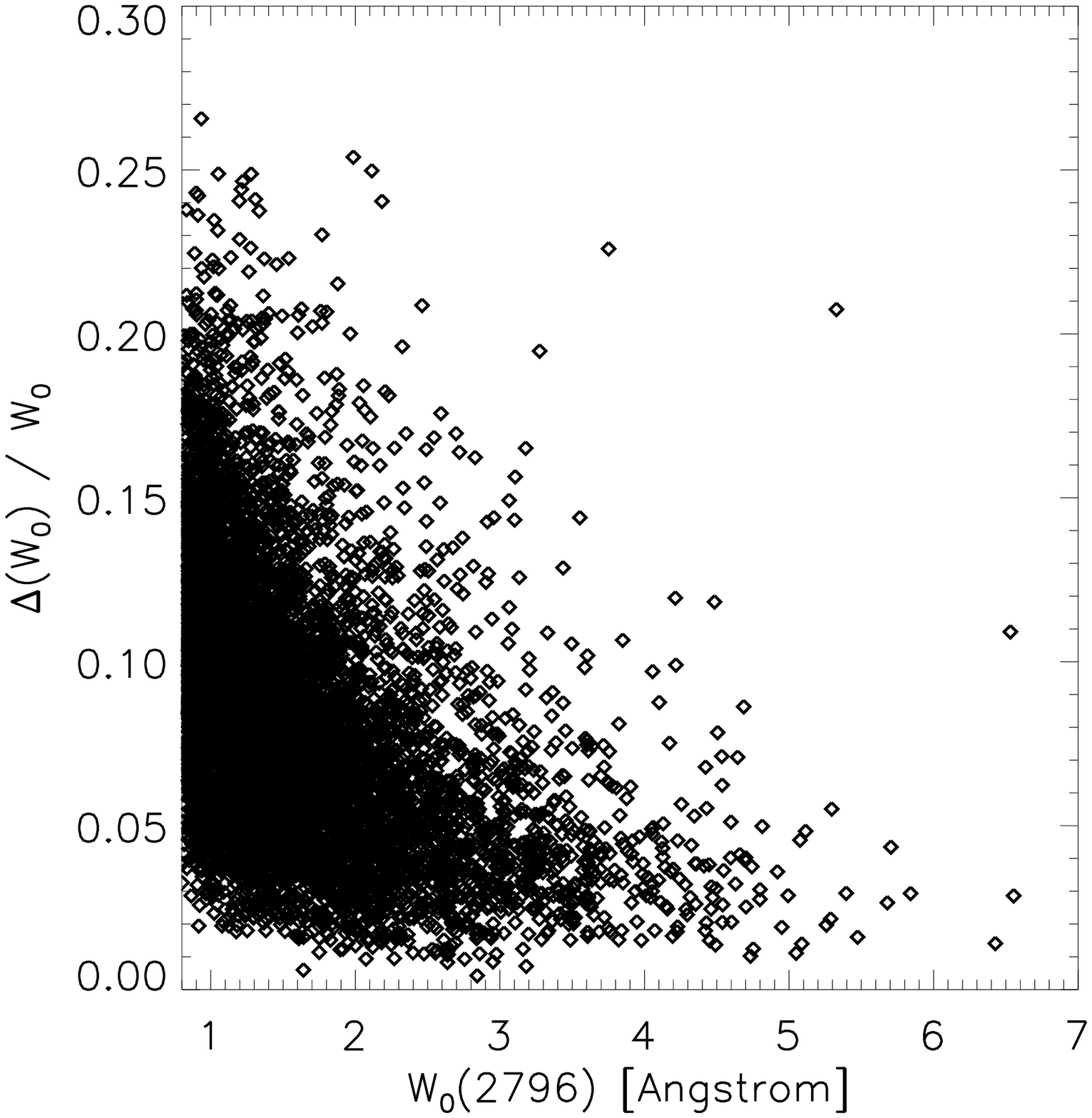}
  \caption{\emph{left panel:} redshift distributions of detected MgII
    absorbers with $W_0>0.8$~\AA\ (filled histogram) and their
    background quasars (solid line). 
    \emph{Middle panel:} rest equivalent width distribution of the
    MgII systems detected in SDSS quasar, \emph{right panel:} observed
    distribution of relative rest equivalent width uncertainty.}
\label{plot_absorber_distribs}
\end{center}
\end{figure*}

Evidence for the presence of dust associated with MgII absorbers has
already been shown by several authors by isolating the reddening and
extinction effects they induce.  Since the selection of quasars
depends on quasar magnitudes and colors \citep{Schneider02}, such an
effect might introduce a bias in the object selection and it is
therefore important to quantify it.  In this section we now
investigate the sensitivity of the SDSS quasar target algorithm
\citep{richards+02} to reddening and extinction \citep[similarly to
the analysis done by ][]{richards+03}.  In order to do so we use
Montecarlo-like simulations to test the ability of the SDSS quasar
algorithm to detect quasars for different amounts of reddening and
extinction.  We proceed as follows:
\begin{enumerate}
\item We use a sample of quasars without strong MgII absorber (as
described in section \ref{section_selection}), consider the SMC
extinction curve and apply various amounts of extinction.  We quantify
the amount of reddening with the $E(B-V)$ excess.
\item The applied extinction will cause some objects to drop out of
the input quasar sample as they become too faint and/or their colors
no longer satisfy the series of selection criteria of the SDSS quasar
selection algorithm, considering only non-radio and point sources.  We
reject the objects no longer labeled as quasars by the SDSS pipeline.
\item For each value of $E(B-V)$ applied, we measure the fraction of
quasars missed due to limiting magnitude and color cuts effects,
we measure the mean $u,g,r,i$ and $z$ magnitudes of the re-observed
quasars and compare them to those of the initial sample of unreddened
quasars.
\end{enumerate}

We illustrate some of the results of this procedure in
Fig. \ref{plot_magnitude_grid}.  From top to bottom we show the
ability of the SDSS quasar selection algorithm to select quasars with
increasing reddening and show the effects on quasar magnitudes. In
each case we present the $u,g,r,i$ and $z$-band magnitude
distributions of the original (solid line histograms) and reddened
samples (filled histograms). These panels show that the fraction of
quasars missed due to dust extinction becomes substantial for observed
reddening values of $E(B-V)$ larger than 0.2.  In this figure we also
show the mean magnitude of each sample with a vertical line. As the
amount of extinction increases, we can easily observe reddening
effects rising as the mean magnitude shifts are more pronounced in
bluer bands. In addition, we can observe that the mean magnitude in
the $z$ band is almost unaffected by reddening in the range considered
here.

We present the summary of the completeness and reddening results in
Figure \ref{plot_Gordon_mag_shift}. From top to bottom we show, as a
function of $E(B-V)$, the input extinction, the measured extinction
and the fraction of objects that can be detected. 
These results show that the observed extinction, defined as the
difference between the mean magnitude of the reddened quasars and
those of the initial sample, is always smaller than the induced
extinction. As explained in section \ref{section_lensing}, such an
effect is expected and is due to the shape of the quasar luminosity
function. Moreover, we can see that the observed mean $z$-band
magnitude is not significantly affected by extinction effects (for the
range of reddening values of interest in this analysis). This property
allows us to use the $z$ band to constrain gravitational lensing
effects.
The bottom panel of the figure allows us to quantify the fraction of
objects missed as a function of observed $E(B-V)$. By knowing the
amount of reddening induced by a class of absorber systems one can use
this curve to infer the fraction of objects missed by the SDSS quasar
finder.

In the right panel of Figure \ref{plot_Gordon_mag_shift}, we show
another property related to the selection of reddened quasars: the
induced and measured \emph{colors} are the same in the range of
reddening values of interest in our case, i.e. with $E(B-V)<0.5$. This
indicates that the fraction of quasars lost due to color cuts is
small. This property therefore allows us to measure the mean color
excess of a sample of quasars and use the corresponding color value to
infer the induced extinction.\\

The properties and selection biases of the SDSS target quasar
algorithm will be used below to interpret the measurements.  Our
analysis is restricted to the main sample of SDSS spectroscopic
quasars (with $i<19.1$) in order to quantify certain selection
effects.  Including fainter quasars is expected to change the above
numbers by a modest amount as the slope of quasar number counts
distribution is a slowly varying function of magnitude (see
\citet{richards+05}).
Redshift-dependent selection effects are not considered here.  As
shown by \citet{richards+02}, the SDSS quasar selection algorithm is
less efficient in detecting reddened quasars at $z\simeq 2.2$. This effect
can be neglected in the present analysis as only a small fraction of objects
are found at such a redshift (see Fig. \ref{plot_absorber_distribs}).

\begin{figure*}[t]
\begin{center}
\includegraphics[height=6cm,width=.3\hsize]{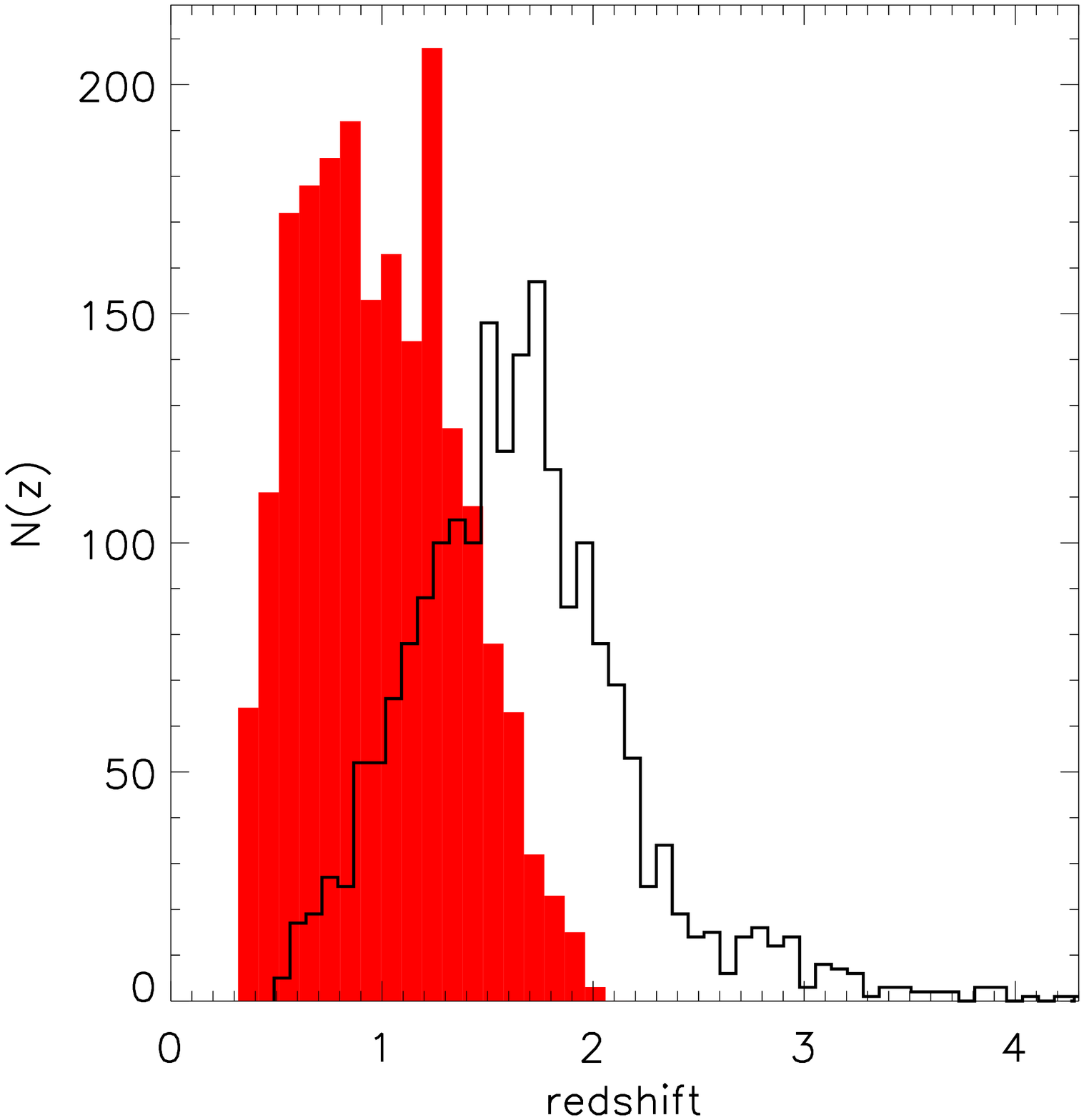}
\includegraphics[height=6cm,width=.3\hsize]{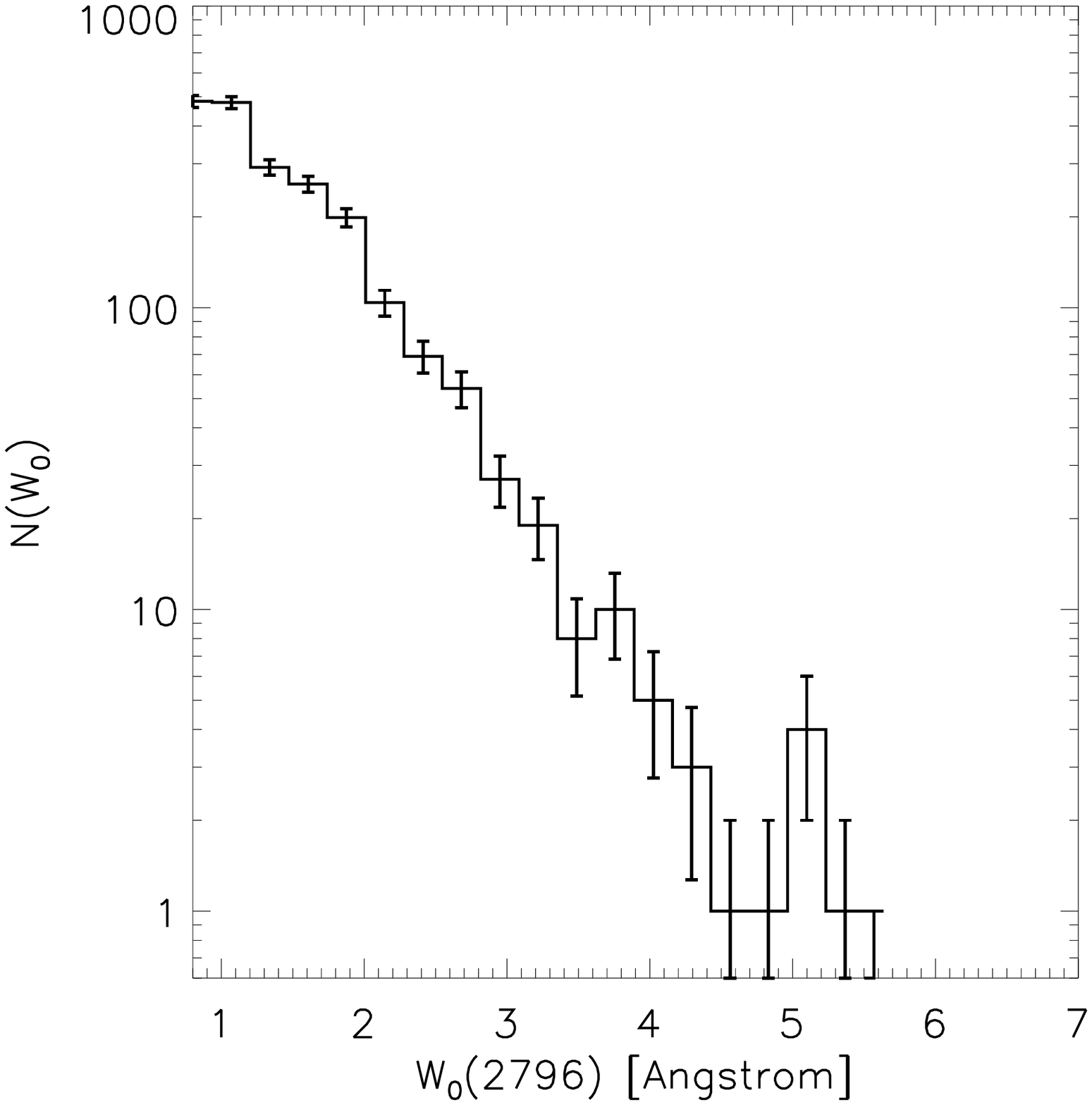}
\includegraphics[height=6cm,width=.3\hsize]{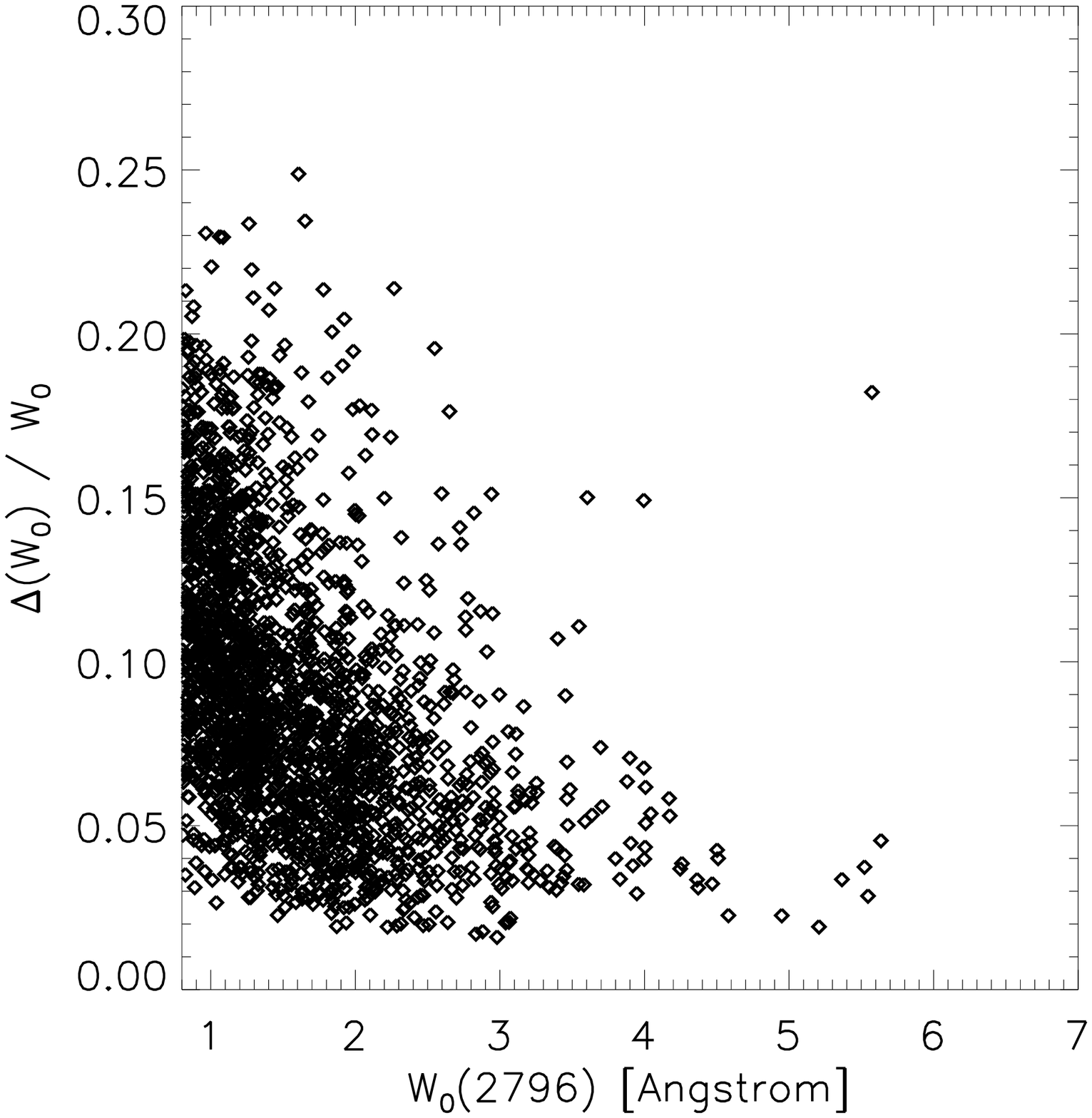}
  \caption{Properties of the simulated absorbers (see figure
  \ref{plot_absorber_distribs} for more information).}
\label{plot_absorber_distribs_fake}
\end{center}
\end{figure*}

\vspace{.5cm}
\subsection{Absorber systems}

\subsubsection{The MgII absorber sample}
\label{section_real}

In order to investigate the effects of reddening and magnification due
to intervening absorbers, we use the sample of \MgII\ systems compiled
with the method presented in \cite{Nestor+05} and based on SDSS DR4
data.  In this section we briefly summarize the main steps involved in
the absorption line detection procedure.  For more details we refer
the reader to \cite{Nestor+05}.

All quasar spectra from the SDSS DR4 database were analyzed,
regardless of QSO magnitude. The continuum-normalized SDSS QSO spectra
were searched for \MgII\ $\lambda\lambda2796,2803$ doublets using an
optimal extraction method employing a Gaussian line-profile to measure
each rest equivalent width $W_0$. All candidates were interactively
checked for false detections, a satisfactory continuum fit, blends
with absorption lines from other systems, and special cases.  The
identification of Mg II doublets required the detection of the 2796
line and at least one additional line, the most convenient being the
2803 doublet partner.  A 5$\sigma$ significance level was required for
all $\lambda2796$ lines, as well as a $3\sigma$ significance level for
the corresponding $\lambda2803$ lines.  Only systems 0.1c blue-ward of
the quasar redshift and red-ward of Ly$-\alpha$ emission were accepted.
For simplicity, systems with separations of less than 500 km/s were
considered as a single system.
The line finder is able to detect MgII absorption lines with a rest
equivalent width $W_0>0.3$~\AA. However a high
completeness is reached only in the range $W_0>0.8$~\AA. We
will therefore focus our analysis on these stronger systems only.
In such a range, multiple absorbers detected in the same quasar
spectrum are found only in $\sim 15\%$ of the systems.  For such cases
we consider only the stronger system.  We note that \MgII\ absorption
lines are in general saturated and in these cases no column density
information can be directly extracted from $W_0^{\lambda2796}$.  The
redshift distribution of the systems used in this analysis is
presented in Figure \ref{plot_absorber_distribs}. The filled histogram
shows that of the MgII systems and the solid line their background
quasars. The middle panel of Figure \ref{plot_absorber_distribs} shows
the \emph{observed} distributions of absorber rest equivalent width
$W_0$. The intrinsic distribution, i.e. corrected for incompleteness
is presented in \cite{Nestor+05}.  The right panel of the figure shows
the distribution of rest equivalent width measurement uncertainty. It
shows the existence of a correlation between line detectability and
absorber rest equivalent width. As can be seen, the relative $W_0$
uncertainty becomes larger for weaker systems.  Because we require the
2796 \AA\ absorption lines to be detected at least at the 5-sigma
level, this results in a selection bias: weaker lines are
preferentially found in high S/N spectra.  In addition, given the fact
that all SDSS quasars are taken with a similar exposure time, the
selection criteria introduce a correlation between quasar brightness
and spectrum S/N. The line finder is therefore expected to
preferentially find weaker systems in brighter quasars.  A crucial
step in our analysis will be to properly take this effect into account
when we define a reference population of quasars without absorbers.

\subsubsection{The simulated MgII samples}
\label{section_simulated}

In order to test the efficiency of our detection technique and to
identify biases and systematic effects we ran Monte Carlo simulations
of the absorber catalog to create a sample of simulated MgII
absorption lines reproducing the same properties as the real sample
($z$, $W_0$, $\Delta W_0$, etc.) and same biases, but with the
difference that the absorption lines are put in random spectra,
i.e. that no correlation exists between quasar magnitude and absorber
properties.  Such a population can then be used as a control sample.

In order to create this sample, we proceed as follows: for each
detected system, a simulated doublet having the same redshift, and
similar $W_0(2796)$, doublet ratio, and Gaussian-fit width was put into many
randomly selected EDR spectra.  Each spectrum containing a simulated
doublet was then run through the entire non-interactive and
interactive pipelines, and $z$ and $W_0(2796)$ were measured for
detected systems. Over 9100 simulated doublets were put in random
quasar spectra. A large fraction of them appear in regions of spectra
with insufficient signal to noise ratio to meet the detection
threshold for the input $W_0(2796)$. The line finder is then able to
recover about 4400 systems.  Measurement error of the rest equivalent
width $W_0(2796)$ can cause systems with $W_0(2796)<W_{lim}$ to
scatter into the output catalog as well as systems with
$W_0(2796)>W_{lim}$ to scatter out.  Therefore, the recovered
distribution of rest equivalent width differs from the input one. In
order to create a simulated absorption line sample with statistically
identical distributions of rest equivalent widths, a trial $W_0(2796)$
distribution was specified and then adjusted so that the simulated
output best represents the data. We initially chose lines randomly from
the input catalog according to a distribution of the form
$\mathrm{N}(W_0(2796))\propto \exp^{-\frac{W_0}{W_{in}}}$ with an initial guess
for $W_{in}$, until the number of lines recovered were equal to that
of the actual catalog. Lines with input $W_0(2796)<0.3$~\AA\ were
used, although only lines with recovered $W_0(2796)>0.3$~\AA\ were
retained.  We determined a maximum-likelihood value for $W_{out}$
using the recovered $W_0(2796)$ values. We then corrected our guess
for $W_{in}$ to minimize $|W_{out}-W_{data}|$.

This procedure allows us to create a simulated absorption line sample
having the same properties as the real sample described in the
previous section.  The distributions of real and simulated absorber
redshifts, rest equivalent widths and rest equivalent widths
uncertainties are presented in Fig \ref{plot_absorber_distribs_fake}.
As can be seen, the statistical properties of the simulated samples
match very well those of the real data.  The figure shows that weaker
absorption lines are preferentially found in high spectrum S/N -- and
therefore brighter quasars -- in the same way as the real systems
are. This bias is systematically introduced by the requirements of the
line finder algorithm. Below we show that the selection of reference
quasars described in the previous section takes this effect into
account.

\subsection{Defining reference quasars}
\label{section_selection}

As our goal is to isolate the magnitude shifts induced by intervening
absorber systems, the selection of reference quasars must (i) not
depend on quasar magnitudes and (ii) take into account the fact that
the absorption line detectability depends on the spectrum S/N as seen
above.

Here we describe how we define such populations of quasars: we first
select the quasars with detected absorbers above a given rest
equivalent width $W_0^{\lambda2796}>0.8$~\AA.  We will compare them
to quasars without any detected absorber above this limit. In the goal
of avoiding biases that would give rise to magnitude shifts, a number
of criteria must be applied: the two quasar populations must have the
same redshift distribution and present the same absorption-line
detectability.  For a given absorber with rest equivalent width $W_0$
and redshift $z_{abs}$, we randomly look for a quasar without any
detected absorber (above $W_0^{\lambda2796}=0.8$~\AA) such that:\\
\indent (i) they have similar redshifts, i.e.
$z_{QSOref}=$$z_{QSOabs}\pm0.1$ ;\\ \indent (ii) in addition, we need
to check whether this quasar would have allowed for the detection of
the absorption line, i.e.\ if its S/N at the corresponding wavelengths
is high enough. Such a requirement translates into
$W_{min}(QSO_{ref},z_{abs})<W_0$, where $W_{min}$ is the minimum
equivalent width that can be detected by the line finder at the
corresponding wavelength.  The quasars selected in this way are called
\emph{reference} quasars in the following.  Our selection procedure
ensures that if similar absorbers population were present in front of
the selected reference quasars, we would have been able to detect
them.\\
Since the number of available quasars without detected
absorbers is larger than the number of quasars with absorbers for the range
of $W_0$ we are interested in, we repeat the previous step $N_{b}$ times
in order to create a series of $N_{b}$ reference samples that will be used
for estimating the noise associated to a single quasar sample.

\section{ANALYSIS \& RESULTS}
\label{section_results}

\begin{figure}[h]
\begin{center}
  \includegraphics[width=\hsize]{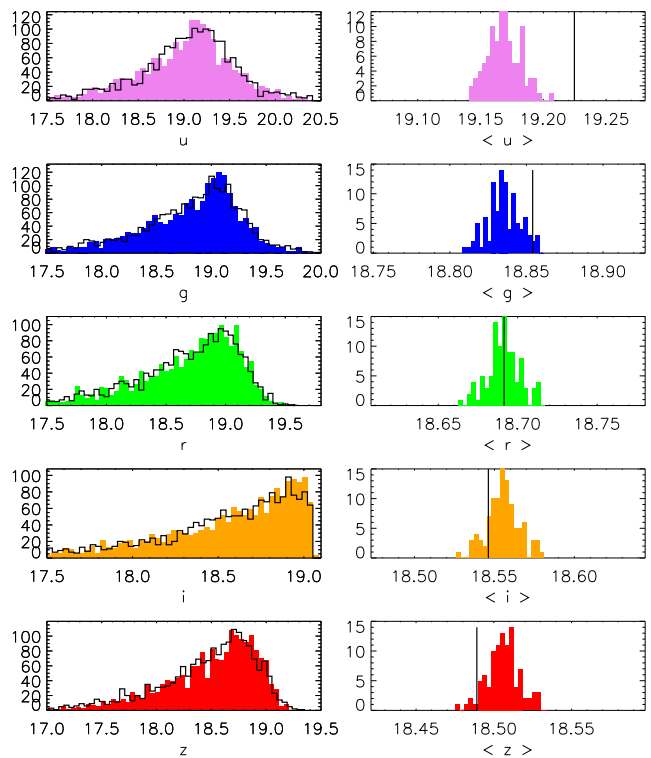}
\caption{The left panels show the magnitude distributions of quasars
with absorbers (solid line histograms) and reference quasars (filled
histograms).  Within the Poisson noise, no significant difference can
be seen between the two types of distributions.  In the right panels,
we show the mean magnitudes of these two populations.  For each band,
the vertical line shows the value
$\left\langle{m_\mathrm{abs}}\right\rangle$ and the histogram shows
the distribution of the mean magnitudes
$\left\langle{m_\mathrm{ref}}\right\rangle$ of $N_{b}$ reference
samples.}
\label{plot_mag_histo}
\end{center}
\end{figure}

In order to isolate the magnitude changes due to the presence of the
absorber systems we measure, in a given band $j$, the mean magnitude
difference between quasars with absorbers and reference quasars:
\begin{equation}
\Delta\left\langle{m_j}\right\rangle=
\left\langle{m_{\mathrm{abs,j}}}\right\rangle-
\left\langle{m_{\mathrm{ref,j}}}\right\rangle\,.
\end{equation}
As explained above, we can repeat the selection of reference quasars
$N_{b}$ times and create a series of reference samples in a
bootstrap manner. The intrinsic dispersion of the series samples can
then be used in order to estimate the errors in the measured magnitude
shifts. 
As an example, we show in Figure \ref{plot_mag_histo} the results
obtained by using a sample of quasars with MgII absorbers having
$1.4<W_0<2$~\AA\ and their corresponding reference quasars. The left
panels show the magnitude distributions of quasars with absorbers
(solid line histograms) and reference quasars (filled histograms).
Within the Poisson noise, no significant difference can be seen
bet\-ween the two types of distributions.  In the right panels, we show
the mean magnitudes of these two populations.  For each band, the
vertical line shows the value
$\left\langle{m_\mathrm{abs}}\right\rangle$ and the histogram shows
the distribution of the mean magnitudes
$\left\langle{m_\mathrm{ref}}\right\rangle$ of $N_{b}=100$ reference
samples. Such an estimator allows to reveal a significant difference
between the magnitude distributions of the two populations and
indicates that quasars with absorbers are affected by reddening
effects.

\begin{figure*}[t]
  \includegraphics[height=7cm,width=.48\hsize]{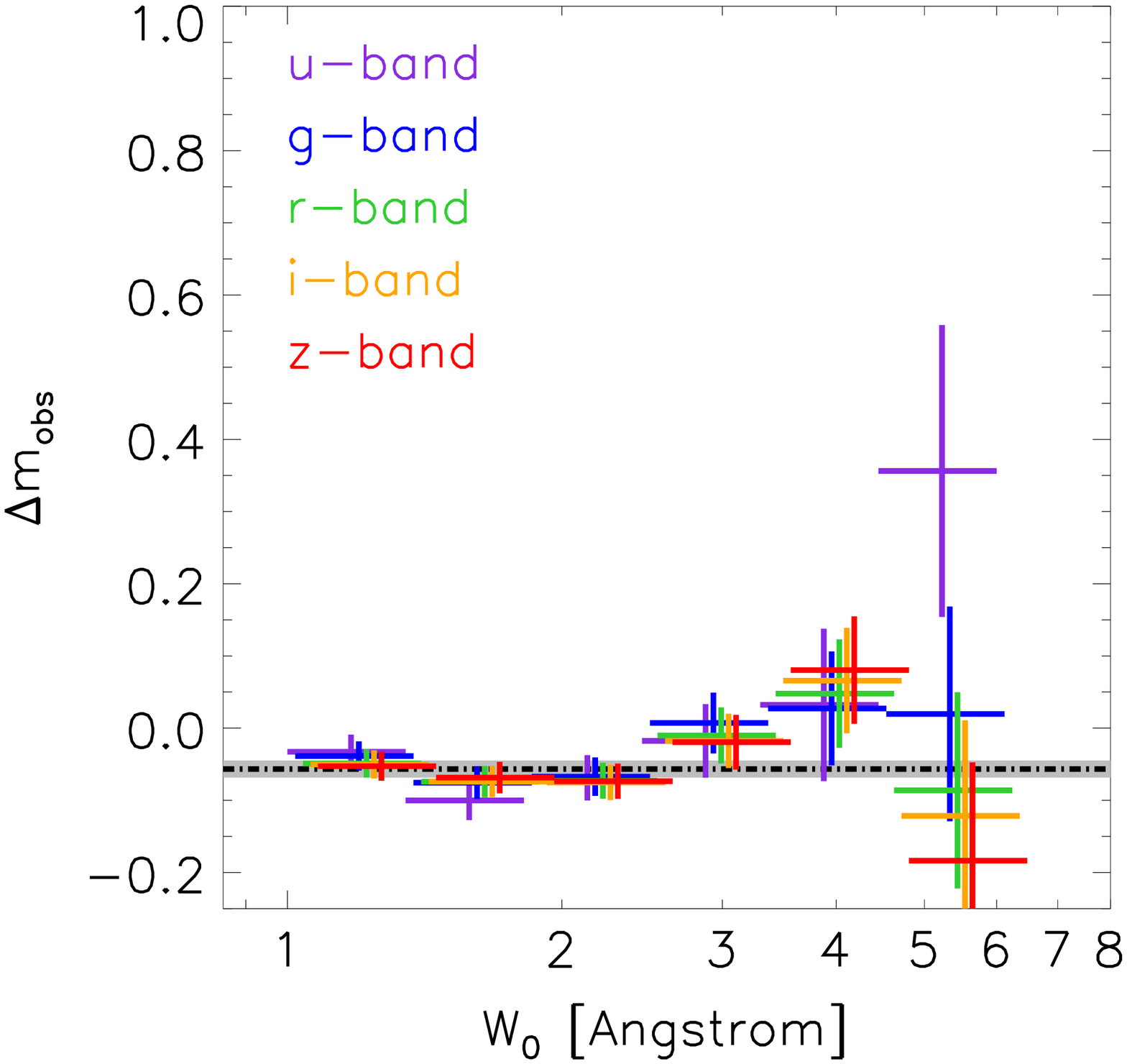}
  \includegraphics[height=7cm,width=.48\hsize]{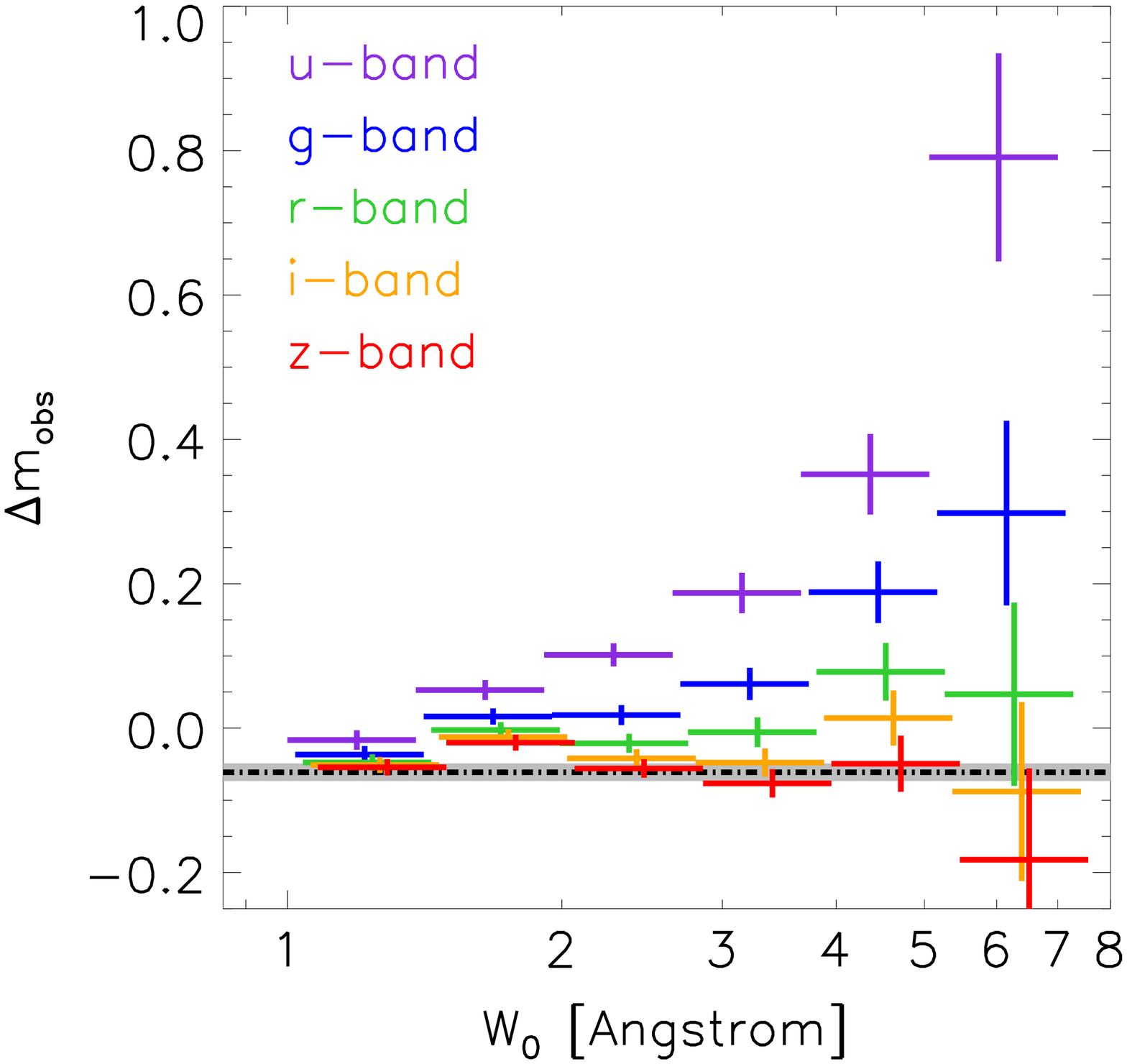}
  \caption{Mean magnitude shifts induced by the presence of absorber
systems, as a function of rest equivalent width.  \emph{Left:} the
results obtained using simulated absorption lines in random quasars
show the existence of a bias giving rise to an excess brightening of
$\Delta m\simeq 0.06$ (indicated by the gray region).  \emph{Right:}
MgII absorbers detected in SDSS quasar spectra induced significant
reddening and extinction effects. When the selection bias is taken
into account, no significant quasar brightening can be observed.}
  \label{plot_magnitude_shift}
\end{figure*}

Before attempting to detect various signals with the data, we make use
of the sample of simulated absorbers in order to test whether our
procedure introduces any selection bias.  Indeed, as measuring the
quantity $\Delta\left\langle{m_j}\right\rangle$ involves the use of
various procedures performed over a large number of objects (continuum
normalization, line detectability, rest equivalent width fitting,
etc.), it is important to investigate the presence of systematic
effects and quantify them.\\
To check whether the properties of quasars with (detected) random
absorption lines match those of quasars without, we proceed as follows:
for a given sample of simulated absorbers, we define a population of
reference quasars as defined in section \ref{section_selection}. We
then compare the mean magnitude difference between the two
populations.  Since the simulated absorption lines were initially put
in random spectra, any departure from zero will quantify the level of
systematics associated to our selection procedure. Such a process can
therefore test whether the samples of quasars with and without
absorption lines to have the statistical properties.
We measure the corresponding mean magnitudes and the results are summarized
in the left panel of Figure \ref{plot_magnitude_shift}.

As can be seen, the mean magnitudes of these two samples are similar,
but a non-zero magnitude difference is found and does not seem to
depend on absorber rest equivalent width or wavelength.  Using the
$z$-band as a reference (the least affected by extinction effects, see
section \ref{section_lensing}), the mean offset is found to be
\begin{equation} 
\Delta\langle m\rangle_{bias}=-0.06\pm0.01 \mathrm{~mag}\,.
\end{equation}
Other bands give similar results, ranging from $-0.05\pm0.01$ to
$-0.06\pm0.01$. This negative value indicates that our global
procedure, i.e. line finder and/or reference quasar selection,
preferentially selects absorbers in brighter quasars. This systematic
effect occurs at the $6\%$ level in terms of quasar magnitude or
similarly spectrum S/N. It could be corrected by increasing the
estimates of $W_{lim}$ by a similar amount. Here we prefer to take
this effect into account by considering the existence of an offset in
the mean magnitude differences.  It should also be pointed out that
our procedure indicates that the selection of absorption lines does
not introduce any color-dependent bias. Indeed, no significant excess
can be detected by comparing the colors of quasars with simulated
absorption lines to those of reference quasars.\\

We now apply the same procedure to the sample of real MgII absorbers.
The results are presented in the right panel of Figure
\ref{plot_magnitude_shift}. The gray region shows the zero point
offset, $\Delta\langle m\rangle_{bias}$, given by the Montecarlo
simulations. As we can see, it appears very clearly that the presence
of a MgII absorber reddens the light of background quasars and the
stronger the absorber rest equivalent width, the redder the quasar.
The direct measurements indicate that $\Delta\langle m \rangle<0$ for
the strongest systems.  However, by taking into account the offset due
to the selection bias $\Delta\langle m\rangle_{bias}$, no significant
difference can be observed. This result contrasts with a number of
previous studies claiming that absorber systems are preferentially
found in the spectrum of brighter quasars. It suggests that previous
claims of such correlations may have arisen due to selection biases.
The current analysis shows that the use of Montecarlo simulations in
order to test the recovery of random absorption lines is an important
task to achieve in order to quantify the amount of systematic effects.
More details on the reddening and magnification effects are presented
below.

\subsection{Dust reddening}
\label{section_dust}

From  the  magnitude shifts  measured  in  each  band we  measure  the
corresponding color excess induced by MgII systems. The measured color
excess  induced  by  the   presence  of  absorbers  provides  us  with
constraints  on  the properties  of  the  dust  associated with  these
systems such  as the size distribution  of the dust  particles and the
related  column  densities.  

\begin{figure}[ht]
\begin{center}
  \includegraphics[height=8cm,width=\hsize]
{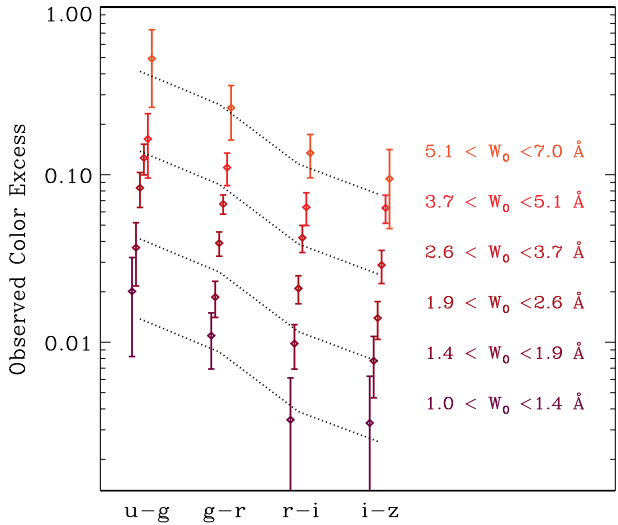}
\caption{Color changes of quasars due to the presence of MgII
    absorbers, as a function of absorber rest equivalent width. The
    solid lines indicate the expected color changes due to an SMC
    reddening curve with different normalizations. The match between
    the slope of this curve and the data points show that the dust
    related to MgII systems is similar to SMC dust.}
  \label{plot_reddening}
\end{center}
\end{figure}

It should be noted that the magnitude shifts measured in different
bands (fig. \ref{plot_magnitude_shift}) are correlated and the errors
on the colors are in general significantly smaller than the errors on
the magnitude shifts. We have therefore computed the color covariance
matrix from bootstrap re-sampling and estimated the errors on the
measured color changes.  The results are presented in
fig. \ref{plot_reddening} where we show the mean color excess measured
for the four color combinations $u-g, g-r, r-i$ and $i-z$, as a
function of absorber rest equivalent width.  As can be seen, strong
MgII absorbers induce, on average, reddening values ranging from
$E(g-r)_{obs}\sim0.02$ to $\sim0.3$, which correspond to similar
$E(B-V)_{obs}$ values.

As the SMC extinction curve is known to describe, on average, the reddening
properties of MgII absorbers (\citealt{Khare+05,Menard+05,York+06}),
we can write
\bea
\big\langle E(B-V) \big\rangle &\simeq& 
\big\langle N_H(z)\,k(z) \big \rangle\,\times\,
\left[ \xi(\lambda_B)-\xi(\lambda_V) \right]
\eea
where $N_H(z)$ is the hydrogen column density associated to a MgII
system, $k(z)$ is the dust to gas ratio, and $\xi(\lambda)$ is the
wavelength dependence of the extinction curve. This relation shows that
investigating the color excess as a function of redshift allows us to
constrain the mean product $\langle N_H(z)\,k(z) \rangle$.
In Figure \ref{plot_reddening} we have plotted the SMC-based reddening
values for a series of amplitudes normalized by the SMC dust column
density
\begin{equation}
\frac{N_d}{N_{d,SMC}}= \left(\frac{k}{k_{SMC}}\right)\,
\left(\frac{N_H}{10^{20}}\right)\,\mathrm{cm}^{-2}\,,
\end{equation}
where $k$ is the dust-to-gas ratio and $N_H$ is the hydrogen column
density.  Our calculation includes the redshift distribution of the
corresponding absorbers in each bin. First, we can see the SMC-like
dust provides a good fit to the measured reddening for most of the
absorber systems.  The dotted lines can be used to quantify the amount
of dust among various strengths of MgII absorbers. They correspond to
$N_d/N_{d,SMC}={1,3,10,30}$ and show that the optical depth for
extinction spans a factor $\sim30$ for MgII systems in the range
$W_0=1$ to $6$~\AA.

If the dust-to-gas ratio of MgII absorbers is comparable of that of
the SMC, it implies that the strongest MgII absorbers have, on
average, hydrogen column densities of
$N_H\sim10^{21}\,\mathrm{cm}^{-2}$, i.e. correspond to the regime of
damped Lyman-$\alpha$ systems. This result is in agreement with the
hydrogen column densities of MgII-selected systems measured by Rao et
al. (2005).
\begin{figure*}[!ht]
\begin{center}
  \includegraphics[width=.45\hsize]{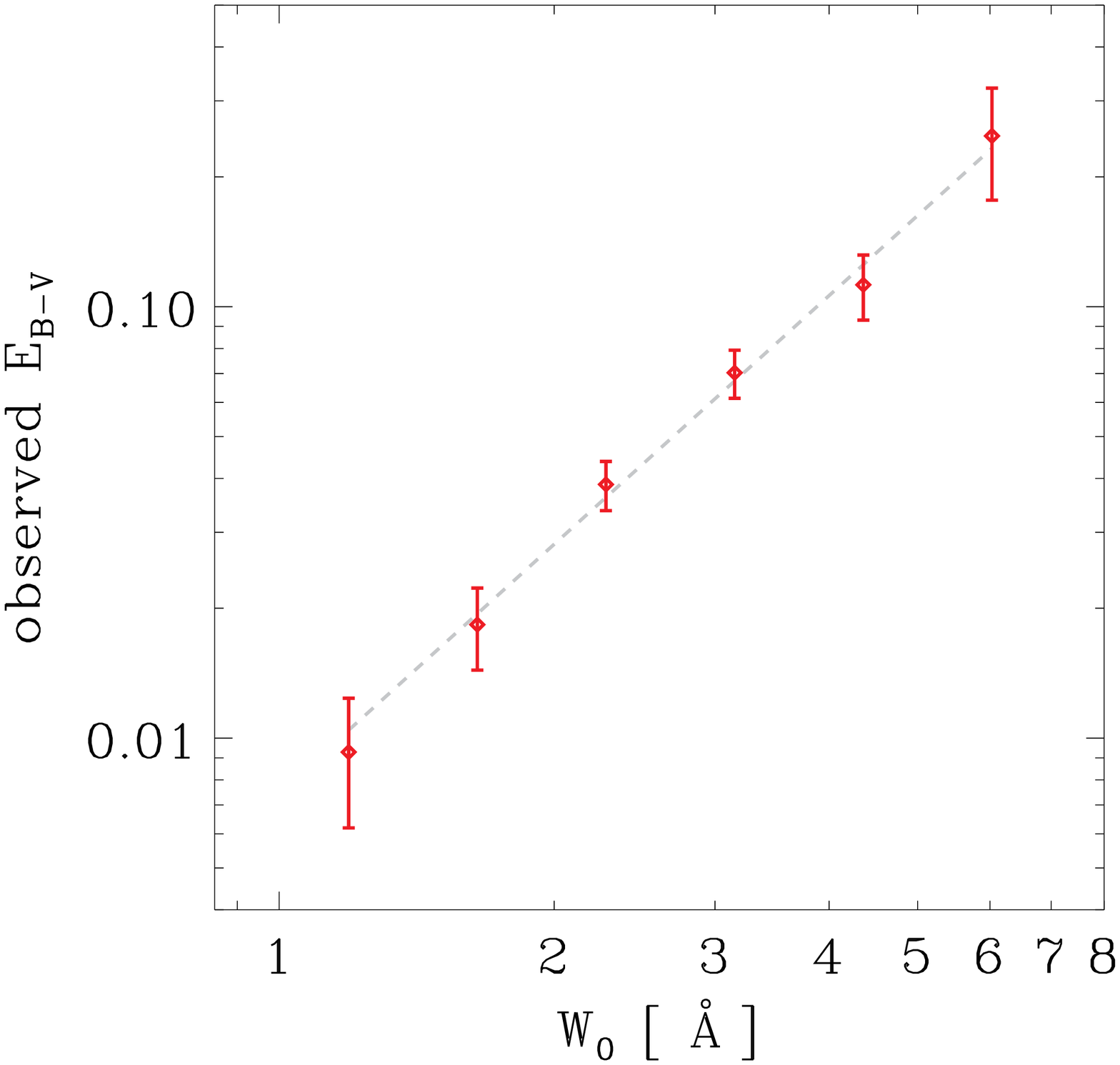}
  \hspace{1cm}
  \includegraphics[width=.45\hsize]{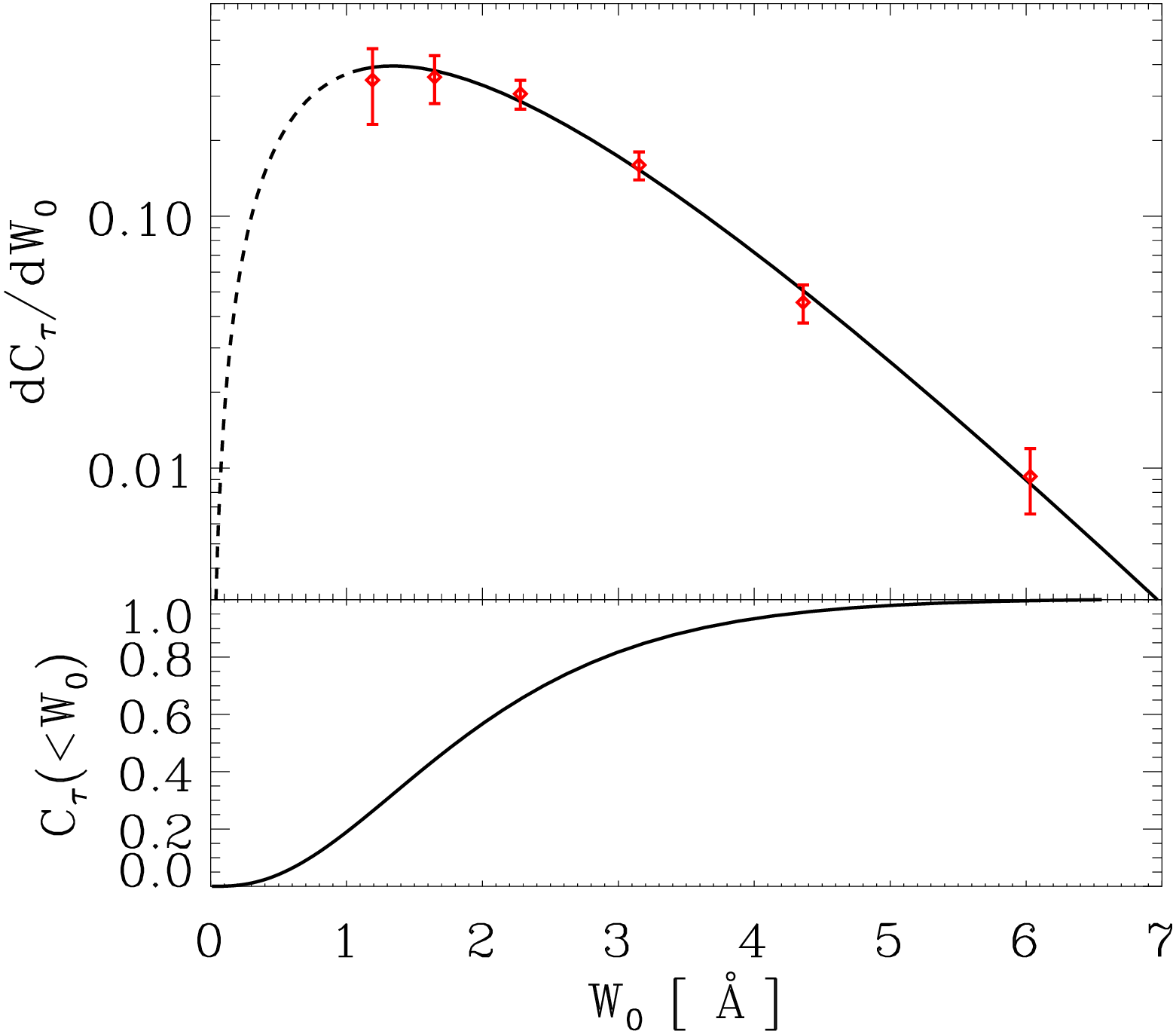}
  \caption{\emph{Left:} Mean E(B-V) values observed as a function of
  absorber rest equivalent width $W_0$. A power-law fit gives:
  $\langle\mathrm{E(B-V)}\rangle=(0.008\pm0.001)\times
  W_0^{1.88\pm0.17}$.  \emph{Right, top panel:} the relative
  contribution of mean dust column density, $\partial C_\tau/\partial
  W_0$, as a function of MgII rest equivalent width. The dashed line
  shows the behavior obtained by extrapolating the observed relation
  between $E(B-V)$ and $W_0$. Using such an assumption, we show the
  cumulative distribution of dust column density $\mathrm{C}(<W_0)$ in
  the bottom panel. Most of the dust is expected to be carried by MgII
  systems with $W_0>1$~\AA.}
  \label{plot_EBV_x_f}
\end{center}
\end{figure*}

As the sensitivity of the SDSS filters is significantly lower in the
$u$ and $z$ bands compared to the $g,r$ and $i$ ones, the $g-i$ color
is measured the most accurately and will be used below to quantify
reddening properties.  We then convert the color excesses $E(g-i)$
into $E(B-V)$ and present the color excesses as a function of absorber
rest equivalent width. The results are presented in
Fig. \ref{plot_EBV_x_f}.  As can be seen, we find a strong
correlation between these two parameters. We can also observe that the
relation between observed mean $E(B-V)$ and $W_0$ is well described by
a simple power-law:
\begin{equation}
\big\langle\mathrm{E(B-V)_{obs}}\big\rangle\,(W_0)=
\rm C \,\left(\frac{W_0}{1\rm\AA}\right)^\alpha\,,
\end{equation}
with $\rm C=(0.8\pm0.1)\times10^{-2}$, $\alpha=1.88\pm0.17$. As shown
in section \ref{section_measured_reddening}, for an SMC-type
extinction law, the color excess $E(B-V)$ is roughly proportionally to
the optical depth for extinction. Hence we can also write
\begin{equation}
\langle \tau_V \rangle (W_0) = \tau_{V,0}\,\left( \frac{W_0}{1\rm\AA} \right)^\alpha\,,
\end{equation}
with $\tau_{V,0}=(2.5\pm0.2)\times 10^{-2}$.

The relation holds for systems with $1<W_0<7$~\AA\ which correspond to
velocity dispersions spanning $100\lesssim\Delta v\lesssim700$ km/s
and the dust column densities change by a factor $\sim30$ in this
range. The weakest and strongest absorbers of this sample are
therefore significantly different but our results indicate
that their dust and gas content follows a universal relation,
suggesting that one phenomenon regulates, on average, the amount of
one quantity with respect to the other.
Modeling the gas distribution in the halo of galaxies is beyond the
scope of this paper but we note that this relation strongly constrains
theoretical models of absorber systems, as the dependence relates
spatial information, $\tau_V$, to a quantity that mostly carries
velocity space information, $W_0$.  We believe that understanding this
relation will greatly improve our knowledge about the nature of these
systems.\\

If we consider the global distribution of strong MgII absorbers it is
interesting to estimate the relative contribution of optical depth for extinction
$\partial C_\tau/\partial W_0$, as a function of absorber rest
equivalent width. Such a quantity is given by the product:
\begin{equation}
\frac{\partial C_{\tau_V}}{\partial W_0} = \frac{\partial N}{\partial W_0}\,\times\,
\big\langle \tau_V \big\rangle (W_0)\,.
\label{eq_tau}
\end{equation}
The number of absorbers per unit rest equivalent width and redshift is
given by Nestor et al. (2005):
\begin{equation}
\partial N/\partial W_0^{\lambda2796} = \frac{N^*}{W^*}
e^{-\frac{W_0}{W^*}}\,,
\end{equation}
with the maximum likelihood values $W^*=0.702 \pm 0.017$~\AA\ and
$N^*=1.187 \pm 0.052$. This estimation takes into account
incompleteness detections and is valid down to a rest equivalent width
limit of $\sim0.3$~\AA. Using this result, we have computed the
relative contribution of dust $\partial C_\tau/\partial W_0$, as a
function of absorber rest equivalent width and present the results
in the upper panel of Figure \ref{plot_EBV_x_f}.  The data points show
that, for MgII systems with $W_0>1$~\AA, most of the dust contribution
originates from systems with $W_0\sim 1-2$~\AA. Therefore most of the
extinction effects are expected to be associated to such systems.  It
is interesting to mention that an excess of MgII absorbers with
$W_0\simeq2$~\AA\ is reported along gamma ray burst lines-of-sight with
respect to those of quasars (Prochter et al. 2006).

We can also attempt to estimate the relative dust contribution
originating from \emph{all} MgII absorbers, i.e. including also weaker
systems.  First, we know that the functional form used for the
incidence of MgII absorbers $\partial N/\partial W_0^{\lambda2796}$
holds down to systems as weak as $\sim0.3$\AA. In the range 0.3 to
1~\AA\ we do not have any detection of dust reddening but we can
attempt to extrapolate the scaling found above (eq. \ref{eq_tau}).  By
doing so, we find that the global dust contribution originating from
systems weaker than 1~\AA\ is smaller than $\sim$20\%.  The result is
similar if, instead of extrapolating eq. \ref{eq_tau}, we use an upper
limit from systems with $W_0\simeq 1$~\AA, i.e. the weakest systems
used in the present analysis.  Therefore, \emph{the vast majority of
the dust probed by MgII absorption lines originates from strong
systems, with $W_0>1$~\AA}.  As dust originates from stars, this
result appears to be in line with the association between strong MgII
absorbers and $\sim L_\star$ galaxies, i.e. galaxies from which most
of the star light originates.

\subsubsection{Dust redshift evolution}
\label{section_redshift}

\begin{figure*}[!ht]
\begin{center}
 \includegraphics[width=.45\hsize]{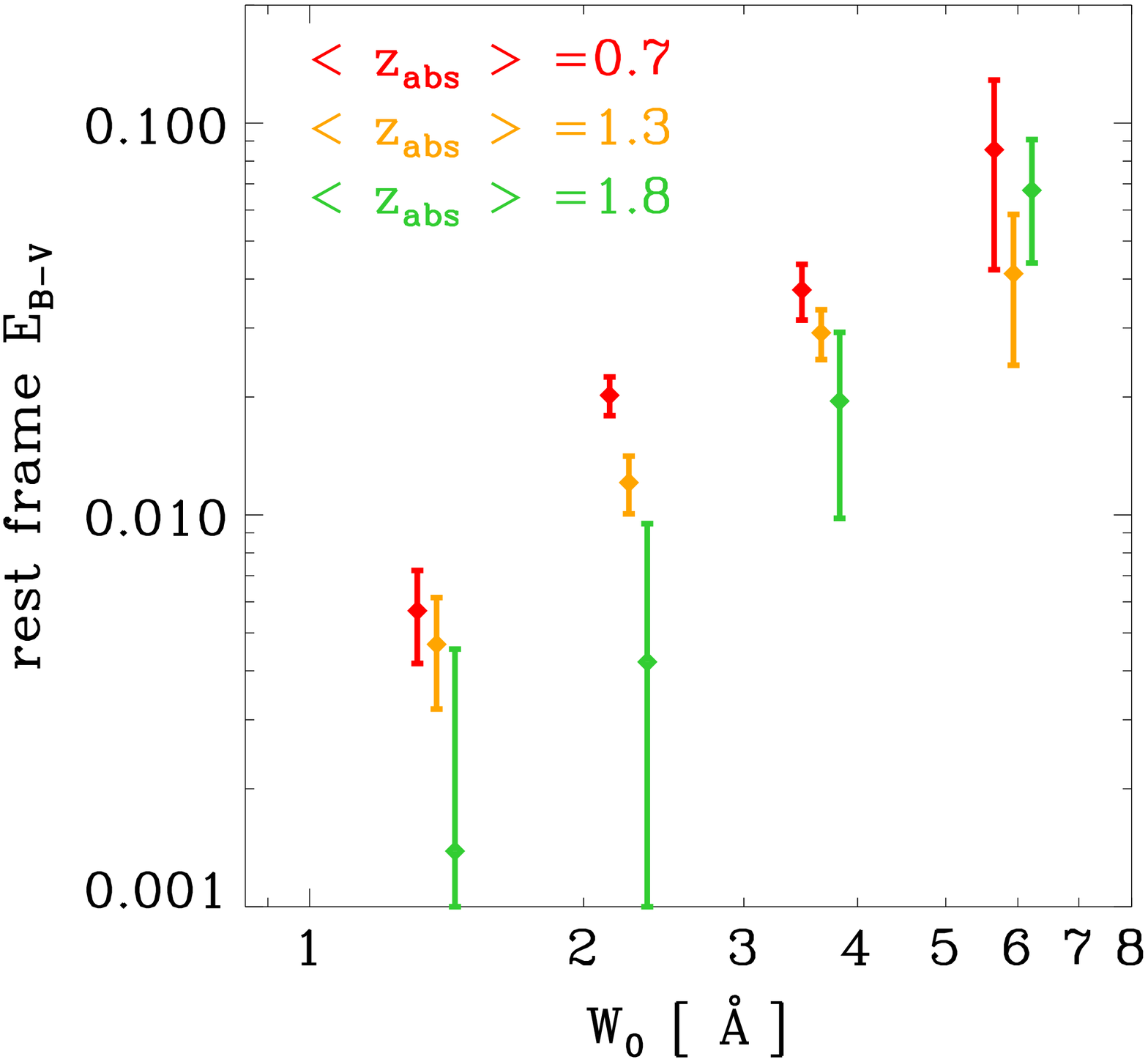}
\hspace{1cm}
  \includegraphics[width=.45\hsize]{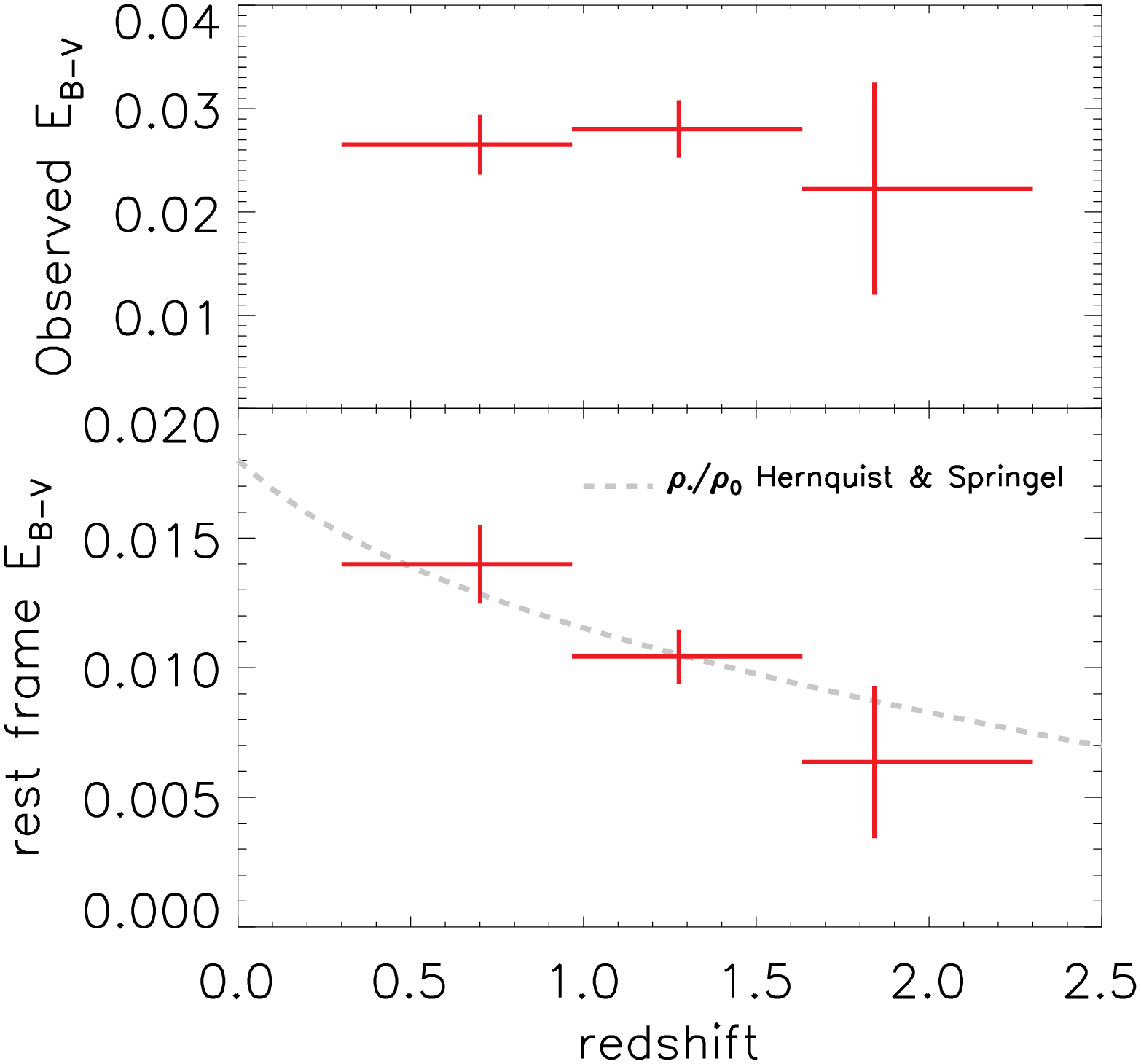}
  \caption{\emph{Left:} Rest frame E(B-V) induced by MgII absorbers
  measured for three redshift bins, and as a function of MgII rest
  equivalent width. \emph{Right:} Evolution of the global E(B-V) (for
  all MgII systems with $W_0>1$~\AA) as a function of redshift. This
  trend shows that the the mean product $\langle N_H(z)\,k(z) \rangle$
  increases by a factor $\sim2$ between $z=2$ and $z=0.5$. This is
  comparable to the evolution of the cosmic star density (or
  integrated star formation rate) illustrated here by the
  (renormalized) model by Hernquist and Springel (2005).}
  \label{plot_ebv_w0_redshift}
\end{center}
\end{figure*}

We now investigate the dust column densities associated with MgII
systems as a function of redshift. Using an SMC-type extinction curve,
we have computed the rest frame E(B-V) values as a function of rest
equivalent width for three redshift bins. The results are presented in
Figure \ref{plot_ebv_w0_redshift} and show that the scaling between
$E(B-V)$ and $W_0$ found in the previous section is consistent with
being redshift independent but the overall amplitude decreases with
redshift indicating that MgII absorbers are less dusty at earlier
times.

In order to quantify the redshift evolution of the dust, we measure
the average $E(B-V)$ of all absorbers systems with $W_0>1$~\AA, in
three redshift bins spanning the range $0.4<z<2.2$. In each case we
measure the observed and rest frame $E(B-V)$ values. The results are
shown in the right panel of Figure \ref{plot_ebv_w0_redshift}. As we
can see, the observed reddening effects of MgII systems are on average
consistent with being redshift independent. Our vision of the distant
universe is equally affected by MgII absorbers at $z\sim0$ and
$z\sim2$.  In contrast, the \emph{rest frame} $E(B-V)$ values, and
therefore the dust column densities, show a significant redshift
evolution.  We find that the amount of dust associated with MgII
absorbers increases by more than a factor two between $z=2$ and
$z=0.4$.  A simple power law fit to the data points gives
$E(B-V)_{rest} \propto (1+z)^{-1.1\pm0.4}$.

It is interesting to see that such a variation is similar to that of
the cosmic star density $\Omega_\star$. We illustrate this agreement
in the figure by showing the integrated star formation rate given by
\citet{hs03} where the scaling has been renormalized to the reddening
values.  Our results therefore provide a new and independent method to
probe the star formation rate over cosmological time scales.

If we assume that the redshift and rest equivalent widths dependencies
can be separated (as suggested by the data points), we can write that
\begin{equation}
\langle E(B-V)_{\rm rest} \rangle\,(W_0,z)=
\rm C' \times \left(\frac{W_0}{1\rm\AA}\right)^\alpha\,(1+z)^\beta\,,
\end{equation}
where $\alpha=1.88\pm0.17$ (see section \ref{section_dust}),
$\beta=-1.1\pm0.4$ and $C'=(0.60\pm0.07)\times10^{-2}$. 
This functional form provides us with interesting constraints on
the properties of cold gas and dust. It will be particularly interesting
to include such information in theoretical models of absorber systems.

\subsubsection{Dust-to-metals ratio}

We can use the above results to estimate the dust-to-metals
ratio of MgII absorbers by computing the ratio
\begin{equation}
{\cal R}_{\rm DM}\equiv\langle E(B-V) \rangle / \langle N(ZnII)\rangle\,,
\end{equation}
adopting Zn (an undepleted Fe-peak element) as an indicator of overall
metal content. This ratio provides us with an estimate ot the fraction
of refractory elements depleted onto dust grains, a quantity
determined by the balance between the formation and destruction of
dust particules. Using a similar sample of MgII absorbers with
$z\sim1$, \citet{Nestor+03} have measured the mean rest equivalent
width of $ZnII\lambda2026$ and found $W_0=0.033\pm0.005$~\AA. Assuming
that most of the zinc is in the form of ZnII, we obtain
$N_{Zn}=(3.63\pm0.5)\times10^{12}$ cm$^{-2}$. Combining this value
with our estimate of the mean E(B-V) of MgII systems at $z\sim1$ we
find
\begin{equation}
{\cal R}_{\rm DM}=(4.1\pm0.6)\times 10^{-15}\, \mathrm{mag\, cm}^2\,.
\end{equation}
As shown by Wild et al. (2006, see their table 5), the dust-to-metals
ratios, ${\cal R}_{\rm DM}$, of the Milky way, LMC and SMC are 4.7,
3.1 and 1.6 $\times 10^{-15}\, \mathrm{mag\, cm}^2$, respectively. Our
analysis therefore shows that, on average, MgII absorbers with
$W_0>1$~\AA\ have a dust-to-metals ratio similar to that of our
Galaxy. The properties of ${\cal R}_{\rm DM}$ as a function of
redshift and MgII rest equivalent width will be presented in a future
paper.

\subsubsection{Dust extinction bias in magnitude limited QSO samples}

\begin{figure}[t]
\begin{center}
  \includegraphics[width=\hsize]{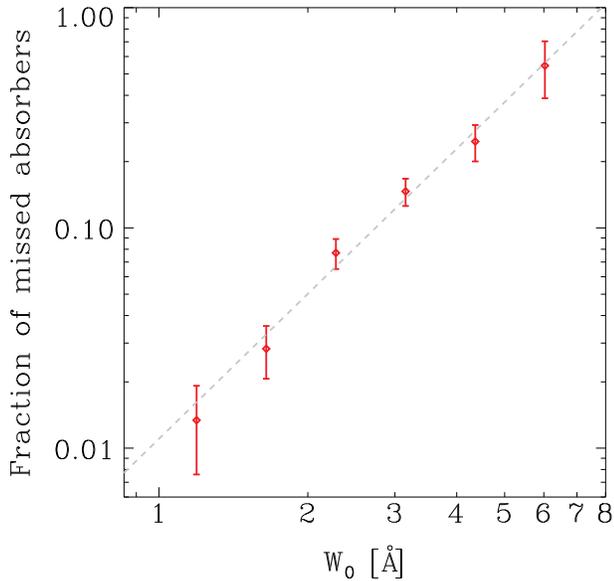}
  \caption{Fraction of quasars with MgII systems that is observed due
to reddening and extinction effects. For the strongest systems we can
see that almost half of the objects are missed in this magnitude
limited sample.}
  \label{plot_missed_fraction}
\end{center}
\end{figure}

From the observed E(B-V) values, we can now use the results of the
reddening Montecarlo simulations presented in section
\ref{section_reddening_bias} and infer the fraction of quasars with
absorbers missed in the magnitude limited sample. Note that our
procedure takes into account quasars missed due to both magnitude and
color selections.
Assuming that the distribution of intrinsic reddening values is not
strongly skewed (which is suggested by the results of \citet{Ellison+04} with
radio-selected quasars), we estimate the fraction of MgII absorbers
missed due to extinction effects and present the results in Fig.
\ref{plot_missed_fraction}. The fraction of missed systems
is below one percent for absorbers with $W_0\simeq1
$~\AA. However, above this value, incompleteness effects become more
important. Our analysis indicates that the SDSS misses more than 10\%
of $W_0\sim 3$~\AA\ MgII absorbers and this rises to more than $50\%$
at $W_0\sim 6$~\AA. By extrapolating the trend to stronger systems, we
see that the SDSS could not detect MgII systems with $W_0\gtrsim
10$~\AA. The fraction of missed quasars with absorbers can be
parametrized by
\begin{equation}
\big\langle{f_m}\big\rangle\,(W_0)=
\rm C_m \,\left(\frac{W_0}{1\rm\AA}\right)^\gamma\,,
\end{equation}
with $C_m=0.005\pm0.001$ and $\gamma=2.23\pm0.18$.\\

The knowledge of $f_m$ allows us to estimate the global fraction of
quasars with strong MgII absorbers, \emph{per unit redshift}, missed
due to the extinction bias. We have
\begin{eqnarray}
f_m^{tot}&=&
\int_{1}^{\infty} f_m(W_0)\,\frac{\partial N}{\partial W_0}\,\d W_0
\nonumber\\
&\simeq&
\int_{0}^{\infty} f_m(W_0)\,\frac{\partial N}{\partial W_0}\,\d W_0
\nonumber\\
&\simeq& 0.01
\end{eqnarray}
Therefore, if we consider quasars at $z=1$, about one percent of them
cannot be observed due to extinction effects by MgII systems.
As both $\partial N/\partial W_0$ and the observed $E(B-V)$ of MgII
systems do not vary strongly with redshift (see section
\ref{section_redshift}), this missing fraction is roughly proportional
to the quasar redshift.  Therefore, if we consider the main sample of
SDSS quasars (for which the mean redshift is above 1.5) the
incompleteness due to the dust hosted by MgII absorbers is about
$2\%$.

\subsubsection{Recovering the unbiased $\partial N/\partial W_0$ distribution}

As the fraction of missed quasars with absorbers depends on $W_0$, the
observed distribution of absorber rest equivalent widths differs from the 
intrinsic one and we have
\begin{equation}
\left( \frac{\partial N}{\partial W_0} \right)_{int} =
\left( \frac{\partial N}{\partial W_0} \right)_{obs} \,
\frac{1}{1-f_m(W_0)}\,.
\label{eq_unbiased_dndW0}
\end{equation} 
As we have used the functional form of $\partial N/\partial W_0$ to
estimate the missing fraction $f_m$, an exact estimation of these
quantities should be made with an iterative process using
eq. \ref{eq_unbiased_dndW0}. However, given the small values of $f_m$
and the exponential cutoff of $\partial N/\partial W_0$, such an
approach will only bring negligible corrections to $f_m$. Our previous
estimations are therefore robust.

The values of $f_m$ shown in Figure \ref{plot_missed_fraction}
indicate that the shape of the observed distribution of $\partial
N/\partial W_0$ is expected to be significantly affected at the
high-end values and to present a deficit.

\subsubsection{Gravitational lensing}
\label{section_lensing}

In order to detect the potential effects of gravitational lensing, we
now estimate the magnitude shifts induced by the presence of strong
MgII absorbers, after correcting for dust extinction.

\begin{figure}[htp]
  \includegraphics[width=\hsize]{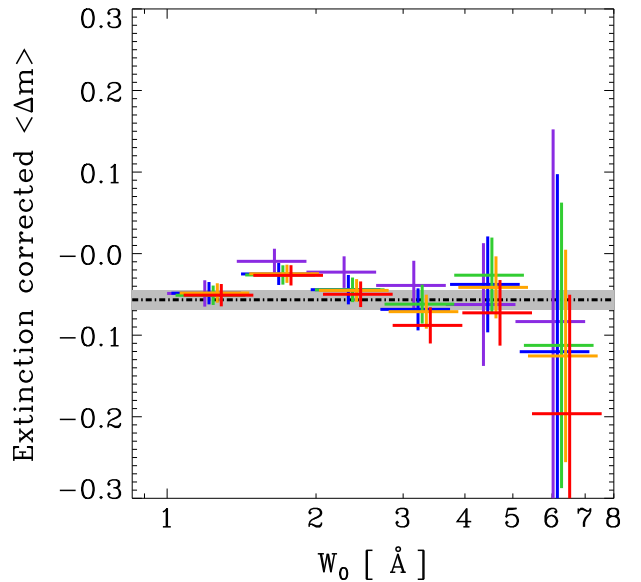}
  \caption{Mean QSO magnitude shift due the presence of a strong MgII
absorber, after correcting for dust extinction, as a function of
absorber rest equivalent width $W_0$. The horizontal line represents
the zero point calibrated with Monte Carlo simulations using simulated
absorption lines in real SDSS quasar spectra. The colors denote the
filter names (see Fig. \ref{plot_magnitude_shift}).  Our analysis does not show any
significant magnification effect, which would correspond to a negative
$\left\langle\Delta m\right\rangle$}
  \label{plot_final_Dm}
\end{figure}

In section \ref{section_reddening_bias} we have shown that, on
average, the induced and measured color excess are equivalent (in the
range of values of interest in our analysis). However, given the shape
of the QSO magnitude distribution and the selection criteria of the
SDSS quasar target algorithm, a difference exists between induced and
measured extinction. Having quantified these effects in section
\ref{section_reddening_bias}, we can now use the observed $E(B-V)$
values and apply the appropriate \emph{observed} extinction correction
to the quasar magnitudes. The corresponding results are shown in Fig.
\ref{plot_final_Dm} where the horizontal line represents the zero
point obtained with the Monte Carlo simulations for simulated
absorption lines. By comparing the extinction-corrected magnitude
shifts to this reference level, we do not find any significant magnification
signal (for which $\Delta m$ would be negative).

If we consider all strong MgII absorbers with $W_0>1$~\AA, we find the
mean magnitude shift in the $z$ band to be:
\begin{eqnarray}
\left\langle\Delta{u}\right\rangle= 0.028 \pm 0.015\;\\
\left\langle\Delta{g}\right\rangle= 0.015 \pm 0.014\;\nonumber\\
\left\langle\Delta{r}\right\rangle= 0.015 \pm 0.013\;\nonumber\\
\left\langle\Delta{i}\right\rangle= 0.015 \pm 0.013\;\nonumber\\
\left\langle\Delta{z}\right\rangle= 0.010 \pm 0.013\;\nonumber
\end{eqnarray}
which was compared to the zero-point and taking into account its
uncertainty. These mean magnitude shifts are consistent with zero
($\left\langle\Delta{u}\right\rangle$ is positive at the 2-$\sigma$
level only). Our analysis does not show any detectable brightening
effect, which differs from a number of previous studies claiming
detections of magnification effects due to metal absorbers. It is
interesting to point out that without the quantification of systematic
effects done with the Montecarlo simulations, i.e. without the
calibration of the zero point, the raw measurement with the real data
would have led to the detection of a negative magnitude shift which
could have been interpreted as magnification effects.\\

We can use the above results to infer an upper limit on magnification effects.
If we consider the $g,r$ and $i$ bands the 3-$\sigma$ upper limit on magnification is
\begin{equation}
\left\langle\Delta m\right\rangle_{\rm lensing} \simeq -0.024\,.
\end{equation}
As explained in section \ref{section_lensing} and detailed in M\'enard
(2005), given the shape of the quasar magnitude distribution, this
corresponds to an upper limit for the mean magnification:
\begin{equation}
\langle \mu \rangle\lesssim 1.10\,,
\end{equation}
meaning that the excess magnification is smaller than about ten
percent.  In order to convert this value into a more interesting
quantity, we can assume the mass distribution of the galaxies
responsible for the absorption to follow that of a singular isothermal
sphere. The galactic halos therefore have the following density and
surface density:
\be
\rho \propto \frac{\sigma_v^2}{r^2}~~~~\mathrm{and}~~~~~
\Sigma(r)=\frac{\sigma_v^2}{2\mathrm{G}r}\;,
\label{eq_sis}
\ee
and the corresponding magnification effects of a point source can then be
computed in a simple way:
\begin{equation}
  \mu = \left\{
  \begin{array}{ll}
    {2/y} & \hbox{if $y\le1$} \\
    1 + {1/y} & \hbox{if $y\ge1$} \\
  \end{array}\right.
\label{eq_mu}
\end{equation}
where $y$ is the impact parameter normalized to the Einstein radius of the
lens: 
\begin{equation}
  \zeta_0 \equiv 4\pi\left(\frac{\sigma_v}{c}\right)^2\,
  \frac{D_{\rm d}\,D_{\rm ds}}{D_{\rm s}}\;,
\label{eq:10}
\end{equation}
where $\sigma_v$ is the velocity dispersion and $D_{\rm d,s,ds}$ are
the angular diameter distances from the observer to the lens, to the
source, and from the lens to the source. Using the above upper limit
on the magnification, and the average impact parameter for MgII
systems with $W_0>1$~\AA\ recently measured by Zibetti et al. (2006):
$\langle b \rangle \simeq 40$ kpc, we find
\begin{equation}
\sigma_v\lesssim269~\rm{km\,s}^{-1}\,,
\end{equation}
for absorbers at $z=0.5$ and quasars at $z=1$. This result shows that,
on average, MgII absorbers are not associated with massive elliptical
galaxies and is in agreement with previous estimates based on
different methods.  Previous works have shown that the average MgII
absorbing galaxy has a luminosity about $\simeq0.8\,L_\star$
\citep{Zibetti+06,Steidel+97}. The expected velocity dispersion of the
gravitational potential of such galaxies is expected to be $\sim 150$
km s$^{-1}$. Therefore the sensitivity of the present measurement
needs to be improved only by a factor of a few in order to detect the
magnification signal originating from the galaxies associated with
MgII absorbers. The present analysis is based only on quasars with
$i<19.1$.  In the future, it might be of interest to quantify the
selection effects of fainter SDSS quasars as it would double the
sample size.  In a longer term, being able to measure this
magnification signal as a function of absorber rest equivalent width
and redshift will provide us with important constraints on the nature
and environment of metal absorbers.

\section{SUMMARY}
\label{section_conclusion}

We have used a sample of almost $7000$ strong MgII absorbers with
$0.4<z<2.2$ detected in the SDSS DR4 dataset to investigate the
gravitational lensing and dust reddening effects they induce on
background quasars.
We have attempted to make significant improvements in making such
measurements, by using a sample of MgII absorbers an order of
magnitude larger than earlier analyzes and by carefully quantifying a
number of systematic effects previously neglected.

In order to do so
we have restricted our analysis to a well defined sample of $\sim
30,000$ SDSS quasars with $i<19.1$ and considered strong MgII
absorbers with a rest equivalent width $W_0>1$~\AA.  We have
quantified the efficiency of the absorption-line finder algorithm with
Monte Carlo simulations and the sensitivity of the SDSS quasar target
algorithm to detect reddened quasars.  We then measured the
statistical magnitude changes induced by the presence of MgII
absorbers. \\

\noindent Our main results are as follows:\\

\textbf{(i)} Strong MgII absorbers significantly redden the light of
their background quasars.  We confirm previous results showing that,
on average, the dust particle size distribution of these systems is
similar to that of the SMC, i.e. without the presence of a reddening
excess at $0.2\micron$.  Moreover, we find that the \emph{observable}
reddening effects follow the simple relation:
$\big\langle\mathrm{E(B-V)_{obs}}\big\rangle\,(W_0)= \rm C
\,(W_0)^\alpha\,,$ with $\rm C=(0.8\pm0.1)\times10^{-2}$,
$\alpha=1.88\pm0.17$, for absorbers with $1<W_0<6$~\AA.  In this
range, the mean dust column density increases by a factor
$\sim30$.  Since most MgII lines with $W_0>1$~\AA\ are saturated, this
scaling suggests that, on average, the dust column density is
proportional to the velocity dispersion of the gas Such a relation
provides us with an important constraint for theoretical models of
absorber systems.
\vspace{.2cm}

\textbf{(ii)} The analysis of reddening effects as a function of has
shown that the \emph{rest-frame} $E(B-V)$ values follows $\langle
E(B-V) \rangle \propto (1+z)^{-\beta}$, with
$\beta=1.1\pm0.4$, which implies that the amount of dust in MgII
systems increases by a factor $\sim2$ between $z=2$ and $z=0.5$. Such
a trend is similar to the evolution of the cosmic star density.  Our
results therefore provide a new and independent method to probe the
star formation rate through the production of dust over cosmological
time scales.\\
The \emph{observed} $E(B-V)$ values induced by MgII appear to be
consistent with no redshift evolution down to $z=2$.  This observation
results from less dust at higher redshift being canceled out by
higher SMC extinction at high redshift (which probe bluer rest-frame
wavelengths).  We have also showed that, considering all MgII
absorbers, most of the dust is carried by systems with
$W_0\sim1-2$~\AA.
\vspace{.2cm}

\textbf{(iii)} We have estimated the dust-to-metals ratio of MgII
absorbers by measuring the quantity ${\cal R}_{\rm DM}\equiv\langle
E(B-V) \rangle / \langle N_{Zn}\rangle =(4.1\pm0.6)\times 10^{-15}\,
\mathrm{mag\, cm}^2$ which is similar to that of the Milky way.
\vspace{.2cm}

\textbf{(iv)} We have quantified the fraction of absorbers and/or quasars
missed due to extinction effects induced by MgII systems. While less
than 2\% of absorbers with $W_0\sim1$~\AA\ are missed, this fraction
increases up to $\sim30\%$ for the strongest systems of our sample,
with $W_0\sim6$~\AA. This effect therefore affects the shape of the
distribution of absorber rest equivalent widths and we provide a
correction factor for it.  We also show that extinction effects due to
MgII systems decrease the number of observable quasars by less than
2\%.
\vspace{.2cm}

\textbf{(v)} Regarding gravitational lensing, in contrast to previous
studies we do not find an excess of absorbers in brighter quasars and
we obtain only an upper limit on gravitational magnification:
$\mathrm{\mu}(W_0^{2796}>1\,\rm\AA)<1.10$. We have shown that testing
the global selection procedure (for detecting absorption lines and
defining reference quasars) is crucial in order to make such a
measurement.  Our result provides an upper limit for the velocity
dispersion of MgII absorbing galaxies: $\sigma_v\lesssim 269$ km
s$^{-1}$.  As it has been shown that the mean luminosity of MgII
absorbing galaxies is about $\sim 0.8 \,L_\star$ \citep{Zibetti+06},
it implies that improving the sensitivity of our measurements by a
factor of a few might lead to the detection of the magnification
effects induced by MgII absorbing galaxies and therefore a characterization
of their mass.\\

\section{Future directions}

Understanding the nature of the structures probed by absorption lines
is an important task to achieved. Among metal lines, the \MgII\
doublet, $\lambda\lambda2796,2803\,$\AA\ is one of the most commonly
detected features in optical quasar spectra but little
is known about the properties and origin of MgII absorber systems.
Attempts to find correlations between $W_0$ and galaxy parameters have
turned out to be rather difficult. Recently, using HST images of 37
galaxies giving rise to MgII absorption, \citet{2007astro.ph..3377K}
showed that most galactic parameters do \emph{not} correlate with
absorptioin strength.

In this paper we have shown the existence of a well-defined scaling
relation between the mean E(B--V) induced by MgII absorbers and their
rest equivalent width $W_0$, and presented its evolution as a function
of redshift.
Since most MgII lines with $W_0>1$~\AA\ are saturated, this scaling
translates into a relation between dust column density and gas
velocity dispersion, with $\Delta v$ spanning the range 100 to about
600 km/s.
Theoretical models are needed in order to gain insight into the origin
of this relation. We believe that being able to understand and
reproduce this scaling will shed light on the nature and properties
of the structures traced by low-ionization absorption lines.

\section*{acknowledgments}

B. M. thanks Nadia Zakamska for her help with using the SDSS pipeline.
Funding for the creation and distribution of the SDSS Archive has been
provided by the Alfred P. Sloan Foundation, the Participating
Institutions, the National Aeronautics and Space Administration, the
National Science Foundation, the U.S. Department of Energy, the
Japanese Monbukagakusho, and the Max Planck Society.

\end{document}